\documentclass{article}

\usepackage[final]{neurips_2025}

\usepackage[utf8]{inputenc} 
\usepackage[T1]{fontenc}    
\usepackage{hyperref}       
\usepackage{url}            
\usepackage{booktabs}       
\usepackage{amsfonts}       
\usepackage{nicefrac}       
\usepackage{microtype}      
\usepackage{xcolor}         
\usepackage{amsmath}
\usepackage{amsthm}
\usepackage{multicol}
\usepackage{multirow}
\usepackage{array}     
\usepackage{graphicx}       
\usepackage{float}
\usepackage{wrapfig}
\usepackage{pifont}
\usepackage{subcaption}
\usepackage{enumitem}
\usepackage{tcolorbox} 
\usepackage{tabularx} 
\newcommand{\xmark}{\textcolor{red}{\ding{55}}}
\newcommand{\cmark}{\textcolor{green}{\ding{51}}}

\usepackage[table,xcdraw]{xcolor}
\usepackage{tcolorbox}
\newtcolorbox{promptbox}[1][]{
  colback=gray!5!white,
  colframe=black!75!white,
  fonttitle=\bfseries,
  title=Prompt,
  #1
}

\title {Cognitive Mirrors: Exploring the Diverse Functional Roles of Attention Heads in LLM Reasoning}
\author{
Xueqi Ma\textsuperscript{1} \quad
Jun Wang\textsuperscript{1,3}\thanks{This work is not related to Amazon.} \quad
Yanbei Jiang\textsuperscript{1} \quad
Sarah Monazam Erfani\textsuperscript{1} \\
\textbf{Tongliang Liu\textsuperscript{2}} \quad
\textbf{James Bailey\textsuperscript{1}} \\
\textsuperscript{1}The University of Melbourne \quad
\textsuperscript{2}The University of Sydney \\
\textsuperscript{3}Amazon \\
\texttt{\{xueqim, jun2, yanbeij\}@student.unimelb.edu.au} \\
\texttt{\{sarah.erfani, baileyj\}@unimelb.edu.au} \\
\texttt{tongliang.liu@sydney.edu.au}
}

\begin{document}

\maketitle

\begin{abstract}

Large language models (LLMs) have achieved state-of-the-art performance in a variety of tasks, but remain largely opaque in terms of their internal mechanisms. Understanding these mechanisms is crucial to improve their reasoning abilities. Drawing inspiration from the interplay between neural processes and human cognition, we propose a novel interpretability framework to systematically analyze the roles and behaviors of attention heads, which are key components of LLMs. We introduce CogQA, a dataset that decomposes complex questions into step-by-step subquestions with a chain-of-thought design, each associated with specific cognitive functions such as retrieval or logical reasoning. By applying a multi-class probing method, we identify the attention heads responsible for these functions. Our analysis across multiple LLM families reveals that attention heads exhibit functional specialization, characterized as cognitive heads. These cognitive heads exhibit several key properties: they are universally sparse, and vary in number and distribution across different cognitive functions, and they display interactive and hierarchical structures.  We further show that cognitive heads play a vital role in reasoning tasks—removing them leads to performance degradation, while augmenting them enhances reasoning accuracy. These insights offer a deeper understanding of LLM reasoning and suggest important implications for model design, training and fine-tuning strategies. The code is available at https://github.com/sihuo-design/CognitiveMirrors.

\end{abstract}

\section{Introduction}

\begin{figure*}[t]
  \centering
  \includegraphics[width=0.9\linewidth]{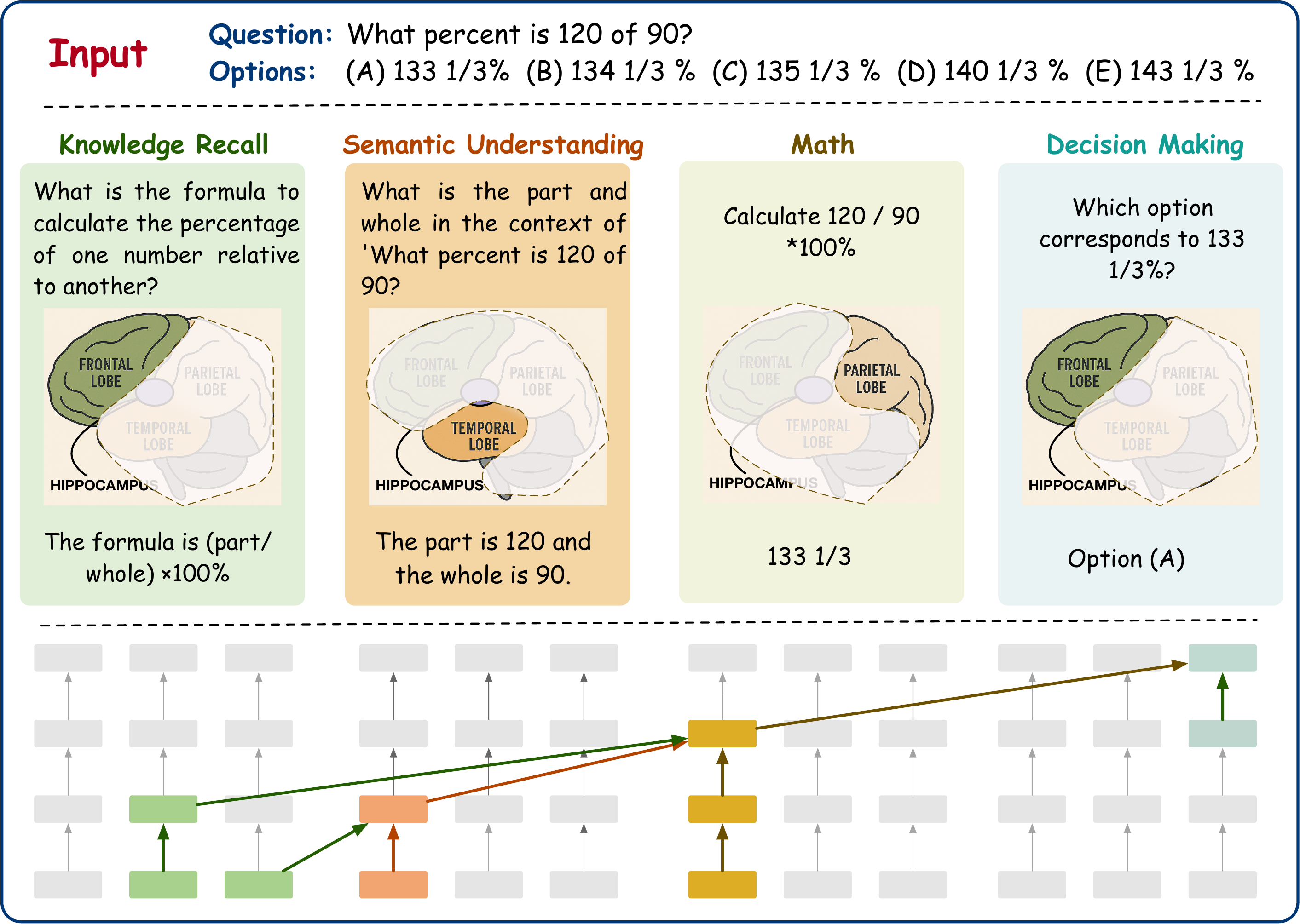} 
  \caption{To solve a complex question, the human brain engages multiple regions to perform distinct cognitive functions necessary for generating a response. We explore whether there are specific attention heads in LLM play functional roles in producing answers.}
  \label{fig:motivation}
\end{figure*}

Large language models (LLMs)~\cite{achiam2023gpt,grattafiori2024llama,touvron2023llama, yang2024qwen2}, built on neural networks that mimic the structure of the human brain, have demonstrated exceptional performance across various natural language processing (NLP) tasks, often exceeding human capabilities. This has sparked growing interest in exploring the potential similarities between the cognitive processes of LLMs and the human brain. Prior studies have demonstrated that LLMs can predict brain responses to natural language~\cite{caucheteux2022deep,schrimpf2021neural}, indicating a functional alignment between artificial models and biological systems. However, to the best of our knowledge, systematic efforts to align reasoning processes between LLMs and human cognitive agents remain scarce.
When solving complex reasoning tasks (e.g., a mathematical multiple-choice question; Figure~\ref{fig:motivation}), the human brain engages a network of specialized regions: the frontal lobe recalls relevant knowledge~\cite{wheeler1997toward}, language areas (e.g., Wernicke’s and Broca’s) support semantic processing~\cite{ono2022bidirectional,meyer2005language}, and the parietal and prefrontal cortices carry out higher-order reasoning~\cite{barsalou2014cognitive,hubbard2005interactions}.

Analogously, recent research suggests that components within LLMs may also take on specialized roles. For example, multi-head attention mechanisms in transformers~\cite{vaswani2017attention} have been found to handle distinct functions, such as information retrieval~\cite{wu2404retrieval} or maintaining answer consistency~\cite{truthful}, pointing toward a form of architectural division of labor. However, most of these findings are based on relatively simple tasks~\cite{zheng2409attention}, leaving open how such specialization operates under complex, multi-step reasoning scenarios.
In parallel, prompting techniques like chain-of-thought (CoT)~\cite{cot} have been shown to improve LLM performance by decomposing complex problems into intermediate steps, a strategy reminiscent of human problem-solving, like the example in Figure~\ref{fig:motivation}. We hypothesize that such prompting may activate and coordinate specialized components within the model. Thus, analyzing the behavior of attention heads under CoT reasoning could contribute insights for a deeper understanding of the internal workings of LLMs and how they process complex tasks.

In this work, we present a novel interpretability framework to systematically analyze the cognitive roles of attention heads during complex reasoning. To facilitate this, we introduce Cognitive Question\&Answering (CogQA), a benchmark dataset that decomposes natural language questions into structured subquestions annotated with fine-grained cognitive functions, such as retrieval, logical inference, and knowledge recall. Leveraging CogQA, we develop a multi-class probing method to identify and characterize attention heads responsible for distinct cognitive operations within the transformer architecture.

We conduct extensive experiments on three major LLM families, including LLaMA \citep{touvron2023llama}, Qwen \citep{yang2024qwen2}, and Yi \citep{young2024yi}. Our results reveal the existence of cognitive heads that consistently exhibit \textbf{universality}, \textbf{sparsity}, and \textbf{layered functional organization} across architectures. Further analysis of the correlations among these cognitive heads reveals clear \textbf{functional clustering}, with heads grouping based on cognitive roles, and uncovers a \textbf{hierarchical structure} in which lower-level heads modulate higher-level ones—mirroring the modular and distributed processing observed in the human cortex \citep{barsalou2014cognitive, ono2022bidirectional}.

Furthermore, we validate the functional importance of these heads by showing that their removal degrades performance on complex tasks and leads to specific error patterns, while their enhancement improves reasoning capabilities. Our findings shed light on the structured cognitive architecture embedded in LLMs and open avenues for function-aware model design and analysis.

\section{CogQA}

In this section, we present a detailed account of our benchmark dataset CogQA’s construction and key characteristics. Although extensive existing benchmark collections span a wide array of NLP tasks, to our knowledge no resource explicitly evaluates LLM reasoning across diverse cognitive functions. To address this, we introduce CogQA, a dataset containing 570 main questions and 3,402 subquestions. Each example comprises a question, its answer, and an annotation specifying the cognitive function required for resolution. 

\subsection{Cognitive Function}

To systematically capture the cognitive processes involved in complex reasoning tasks, we categorize cognitive functions into two groups: low-level functions and high-order functions, inspired by established frameworks in cognitive science~\cite{anderson2014rules,diamond2013executive}. Low-level functions primarily involve information retrieval and linguistic analysis, while high-order functions engage more abstract reasoning, problem-solving, and decision-making. \textcolor{black}{Detailed descriptions of these cognitive functions are provided in Appendix \ref{app:prompt_generating}.}

The low-level cognitive functions include:
\begin{itemize}[noitemsep,topsep=0pt]
    \item \textbf{Retrieval}: locating relevant information from an external source or prior context.
    \item \textbf{Knowledge Recall}: accessing stored factual or procedural knowledge from memory.
    \item \textbf{Semantic Understanding}: interpreting the meaning of words, phrases, or concepts.
    \item \textbf{Syntactic Understanding}: analyzing the grammatical structure of a sentence.
\end{itemize}

The high-order cognitive functions include:
\begin{itemize}[noitemsep,topsep=0pt]
\item \textbf{Mathematical Calculation}: performing arithmetic or numerical operations.
\item \textbf{Logical Reasoning}: drawing conclusions based on formal logical relationships.
\item \textbf{Inference}: deriving implicit information that is not directly stated.
\item \textbf{Decision-Making}: selecting the best outcome among alternatives based on reasoning.
\end{itemize}

This categorization reflects a natural progression from basic information processing to complex cognitive integration.  Both the human brain and LLMs encompass a wide range of functional modules. Our focus in this work is specifically on reasoning-related cognitive functions. By identifying and organizing these eight core reasoning functions, we can more clearly examine how LLMs handle different types of thinking steps, in a way that is both systematic and easy to interpret.

\subsection{Data Collections}

Based on our categorization of cognitive functions, we sampled 750 diverse questions from NLP reasoning benchmarks, selecting 150 examples from each of AQuA~\cite{aqua}, CREAK~\cite{creak}, ECQA~\cite{ecqa}, e-SNLI~\cite{esnli}, and GSM8K~\cite{gsm8k}. These datasets cover a range of reasoning types, including logical, mathematical, and commonsense reasoning. Using the CoT paradigm, we prompted GPT-4o~\cite{hurst2024gpt} to decompose each question into subquestions, each targeting a single cognitive function. The prompt encourages structured, step-by-step reasoning, with each subquestion being clear, answerable, and sequentially dependent. This yields a set of subquestion-answer-cognitive function (subQAC) triples for each QA pair:  $\operatorname{subQACs}=\left\{\left(q_i, a_i, c_i\right)\right\}_{i=1}^k$, where each contains a subquestion $q_i$, its concise answer $a_i$, and the corresponding cognitive function label $c_i$. The prompt for generating subquestions and examples are list in Appendix~\ref{app:prompt_generating} and Appendix~\ref{app:exmaple}, respectively.


\subsection{Data Filtering and Annotation}

Recent advances have made it increasingly feasible to use LLMs for dataset construction, owing to their strong reasoning abilities and capacity to generate high-quality annotations at scale~\cite{llm_annotate}. Although our dataset is constructed automatically using an LLM to reduce manual effort, we implement a strict two-stage human verification pipeline to ensure data quality and mitigate hallucinations. In the first stage, three expert annotators independently assess whether the subquestions are logically structured and align with natural human reasoning. QA pairs with inconsistent or incoherent decompositions are filtered out. In the second stage, annotators verify and, if necessary, relabel the cognitive function associated with each subquestion to ensure alignment with the intended mental process. Finally, we validate the subanswers by cross-checking them using the GPT-o4-mini model~\cite{o4mini2024}, followed by human adjudication where discrepancies arise. Details of the annotation process and rubric can be found in Appendix~\ref{app:annotation}.
This multi-step filtering ensures that each retained subQAC triple reflects a coherent, interpretable reasoning step grounded in core cognitive functions. After this refinement, our final dataset contains 570 main QA and 3,402 validated subQAC triplets.

\section{Cognitive Function Detections}

Given the CogQA dataset, we aim to identify which attention heads in LLMs are associated with specific cognitive functions. We adopt a probing-based framework, a widely used interpretability technique in which an auxiliary classifier is trained to predict properties from intermediate model representations~\cite{alain2016understanding, belinkov2022probing, tenney2019bert}. We frame this as a multi-class classification task: for each cognitively annotated subquestion, we extract head activations (see Section~\ref{extract}), train classifier and compute importance scores to identify contributing heads (see Section~\ref{importance}). Unlike prior work focusing on a single-class, our method captures many-to-many relationships between heads and functions, enabling a more detailed analysis of functional specialization and overlap compared to prior single-class approaches.

\subsection{Head Feature Extraction}\label{extract}

Given a large language model $\mathcal{M}$, we generate an answer $a_i^{\mathcal{M}}$ for each subquestion $q_i$ derived from a main question $Q_i$. To support coherent multi-step reasoning, we include preceding subquestions and their answers as contextual input, emulating the incremental reasoning process observed in human cognition.

During inference, input tokens are embedded and processed through successive transformer layers. At each layer, attention and feedforward operations update the residual stream, which is ultimately decoded into token predictions. For each generated token $i$, we extract attention head outputs $X_i = \{ x_l^m(i) \mid l = 1, \dots, L,\ m = 1, \dots, M \}$ across all layers, where $x^m_l$ denotes the value vector from the $m$-th head in layer $l$ projected into the residual stream, with $M$ the number of heads per layer and $L$ the total number of layers.

Let $N_t$ denote the number of tokens in the generated answer $a_i^{\mathcal{M}}$. To isolate semantically informative content relevant to reasoning, we select the top-$k$ most important tokens,~\footnote{We include an ablation study in Appendix~\ref{app:abl} to analyze the impact of using alternative token positions.} determined by prompting GPT-o4-mini \cite{o4mini2024} (skilled in reasoning), yielding an index set $\mathcal{I}_k$ with $|\mathcal{I}_k| = k$ (Top-$k$ ($k=5$) token examples are in Appendix \ref{app:topk}). For each index $j \in \mathcal{I}_k$, we extract the corresponding attention head activations $X_j$, and compute the averaged activation feature for the $m$-th head in layer $l$ as $\bar{x}_l^m = \frac{1}{k} \sum_{j \in \mathcal{I}_k} x_l^m(j)$. This results in a full set of head-level features $\bar{X} = \{\bar{x}_l^m \mid l = {1, \ldots, L},\ m = {1, \ldots, M}\}$. 

Given prior findings suggesting that cognitive functions may vary by layer depth~\cite{zheng2409attention}, we incorporate layer-wise information by computing the average activation $\bar{x}_l=\frac{1}{M} \sum_{m=1}^M \bar{x}_l^m$ for each layer. We then augment each head-level vector with its corresponding layer summary, resulting in enriched features $\bar{x}^{m'}_l = [\bar{x}^m_l;\bar{x}_l]$. For each subQA triplet $(q_i,\ a_i,\ c_i)$, the final input to the probing classifier is given by $\{\bar{x}^{m'}_l \mid l = {1, \ldots, L},\ m = {1, \ldots, M}\}$.

\subsection{Heads Importance}\label{importance}

For the CogQA dataset with $N$ subQA pairs, we collect all activations to construct the probing dataset:

\begin{equation}
    \mathcal{D}_{\text{probe}} = \left\{ (\bar{x}^{m'}_l,\ c)_i \right\}_{i=1}^{N}, l \in \{1, \ldots, L\},\ m \in \{1, \ldots, M\}
\end{equation}


We split the dataset into training and validation sets with a $4{:}1$ ratio. Each attention head feature is first passed through a trainable linear projection for dimensionality reduction, followed by a two-layer MLP that performs multi-class classification over cognitive functions (training details are provided in Appendix~\ref{mlp}). To interpret the contribution of individual heads to each function, we use a gradient-based attribution method. Specifically, for each function class \(c\), we compute the contribution of each head feature via the gradient$\times$activation technique:

\begin{equation}
    I^{(c)}_j = \mathbb{E}_{(\bar{x}, c) \sim \mathcal{D}_{\text{probe}}} \left[ \frac{\partial \hat{y}_c}{\partial \bar{x}_j} \cdot \bar{x}_j \right],
\end{equation}
where \(\bar{x}_j\) is the \(j\)-th head input feature, and \(\hat{y}_c\) is the classifier’s predicted logit for class \(c\). This yields an importance score for each attention head with respect to each cognitive function. We aggregate the scores into a matrix \(\mathbf{I} \in \mathbb{R}^{C \times (L \cdot M)}\), where each row corresponds to a function class and each column to a specific head in a specific layer.

We hypothesize that attention heads with higher importance scores contribute more significantly to each cognitive function. By ranking heads according to their importance, we can identify which heads and layers are specialized for specific functions. Subsequent targeted interventions on these heads validate the effectiveness of this approach.




\section{Experiments}

We conduct a series of experiments on three LLM families across various model scales, including LLaMA \cite{touvron2023llama} (Llama3.1-8B-instruct and Llama3.2-3B-instruct), Qwen \cite{yang2024qwen2} (Qwen3-8B and Qwen3-4B), and Yi \cite{young2024yi} (Yi1.5-9B and Yi1.5-6B). Our goal is to identify cognitive attention heads associated with specific reasoning functions and evaluate their roles via targeted interventions. By selectively masking these heads, we assess their functional significance in supporting downstream performance. We evaluate our method in terms of functional alignment, consistency across models, and causal impact on reasoning tasks. Results confirm the existence of sparse, function-specific heads and highlight their critical contribution to structured cognitive processing within LLMs.



\subsection{Properties of Cognitive Heads}

\begin{figure}
  \centering
  \includegraphics[width=1\linewidth]{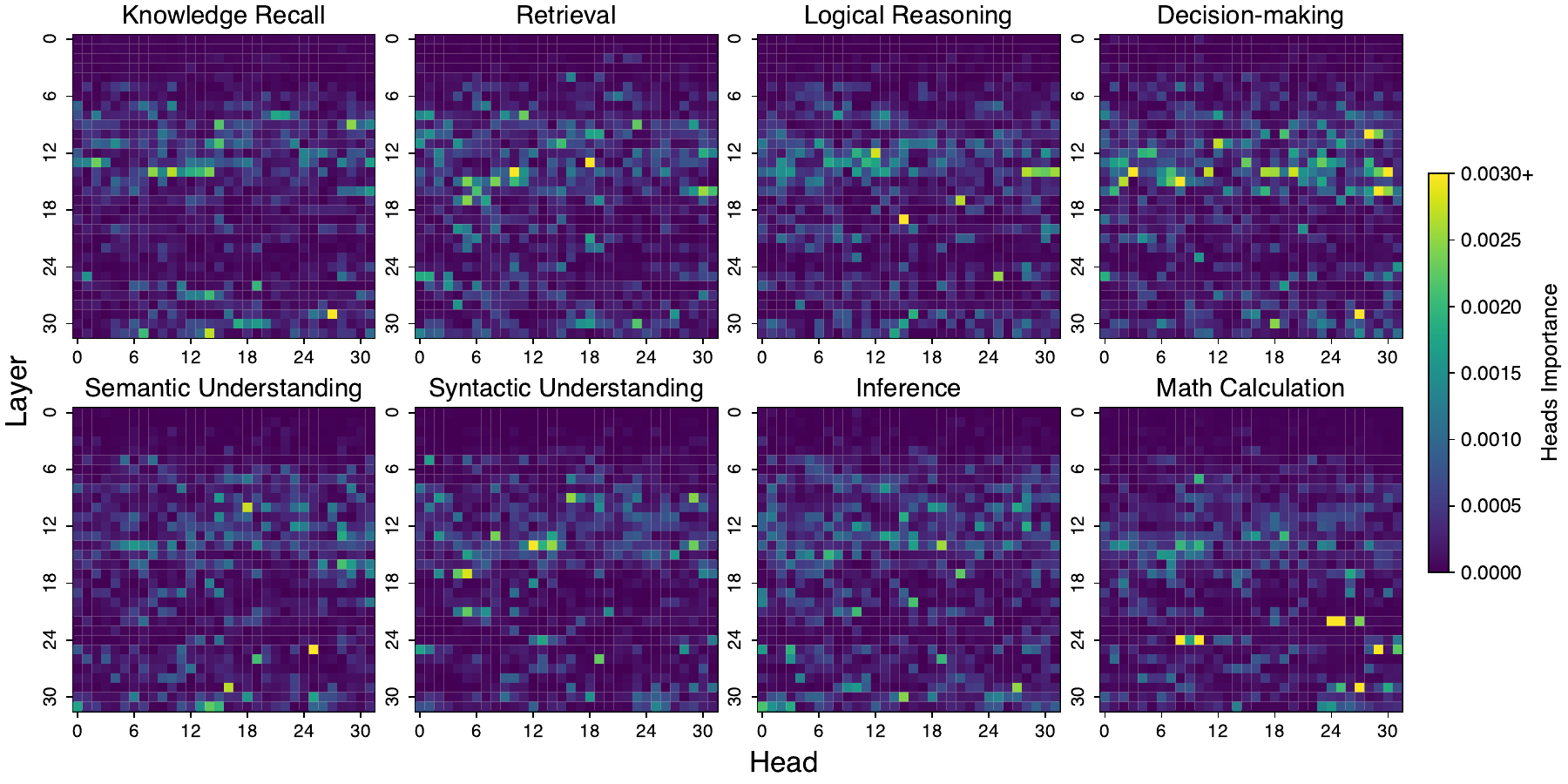} 
  \caption{The existence of cognitive heads in Llama3.1-8B-instruct responsible for eight distinct functions in complex reasoning tasks. The x-axis represents the head index, while the y-axis indicates the layer index.}
  \label{fig:heatmaps}
\end{figure}

Our analysis reveals that cognitive head importance in large language models exhibits three key properties: \textbf{sparsity and universality}, and \textbf{layered functional organization}. To illustrate these characteristics, we present the heatmap of attention head importance scores across eight cognitive functions in Llama3.1-8B-instruct (Figure~\ref{fig:heatmaps}). 

\textbf{Sparsity and Universality:} As shown in Figure~\ref{fig:heatmaps}, each cognitive function activates only a small number of high-importance attention heads, revealing a strikingly sparse pattern. In Llama3.1-8B-instruct, fewer than 7\% of all heads have importance scores above 0.001 across the eight functions, suggesting that only a compact subset of heads meaningfully contribute to task performance. This sparsity is not uniform: Retrieval contains the highest proportion of salient heads (6.45\% exceeding 0.01), while Inference has the fewest (3.42\%). These results highlight that LLMs rely on highly specialized, localized components for different cognitive abilities. Importantly, we observe that this sparse functional organization is consistent across different model architectures and sizes. Additional heatmaps for five other models are provided in Appendix~\ref{app-heatmaps}, supporting the universality of this phenomenon.

\textbf{Layered Functional Organization:} In addition to sparsity, attention heads show a structured distribution across model layers. Retrieval-related heads cluster primarily in the middle layers, while math-related heads appear more frequently in higher layers. This structured, task-dependent localization points to an emergent modular organization, where different layers support distinct cognitive operations. Further, we identify cognitive heads by selecting those before the elbow point of each function’s descending importance curve (Appendix~\ref{app-curve}), and find notable variation in head counts across functions (Appendix~\ref{app-number}). For example, in the LLaMA family, mathematical calculation requires fewer heads (59 in Llama3.1-8B-Instruct, 35 in Llama3.2-3B-Instruct), while inference draws on substantially more (139 and 98, respectively), reflecting differences in representational and computational complexity.

\begin{table*}[t]
\caption{Intervention results (\%) of cognitive heads vs. random heads across 8 cognitive functions: \textbf{Retrieval}, Knowledge \textbf{Recall}, \textbf{Semantic} Understanding, \textbf{Syntax} Understanding, \textbf{Math} Calculation, Inference, \textbf{Logic} Reasoning, and \textbf{Decision} Making. Lower values indicate more effective intervention outcomes, suggesting that the corresponding heads play a greater role in the cognitive function.}
    \centering
    \setlength{\tabcolsep}{3pt}
    \resizebox{1\linewidth}{!}{
    \renewcommand{\arraystretch}{1.0}
    \begin{tabular}{lc|cccccccccccccccc}
        \toprule
        \multirow{3}{*}{Model} & \multirow{3}{*}{Inter\_Head} 
        & \multicolumn{8}{c}{Information Extraction and Analysis Functions} 
        & \multicolumn{8}{c}{Higher-Order Processing Functions} \\
        \cmidrule(r){3-10} \cmidrule(r){11-18}
        & & \multicolumn{2}{c}{Retrieval} & \multicolumn{2}{c}{Recall} 
        & \multicolumn{2}{c}{Semantic} & \multicolumn{2}{c}{Syntactic}
        & \multicolumn{2}{c}{Math} & \multicolumn{2}{c}{Inference} 
        & \multicolumn{2}{c}{Logic} & \multicolumn{2}{c}{Decision} \\
        \cmidrule(r){3-4} \cmidrule(r){5-6} \cmidrule(r){7-8} \cmidrule(r){9-10}
        \cmidrule(r){11-12} \cmidrule(r){13-14} \cmidrule(r){15-16} \cmidrule(r){17-18}
        & & comet & acc & comet & acc & comet & acc & comet & acc 
        & comet & acc & comet & acc & comet & acc & comet & acc \\
        \midrule

        \multirow{2}{*}{Llama3.1-8B} 
        & random & 90.83 & 84.71 & 87.85 & 83.84 & 91.44 & 97.50 & 87.81 & 66.17 & 94.25 & 83.08 & 91.90 & 70.18 & 91.39 & 54.69 & 97.64 & 90.91 \\
        & \textbf{cognitive} & \textbf{44.96} & \textbf{8.24} & \textbf{56.93} & \textbf{38.38} & \textbf{81.98} & \textbf{75.00} & \textbf{69.20} & \textbf{40.00} & \textbf{87.81} & \textbf{66.17} & \textbf{76.65} & \textbf{52.63} & \textbf{52.07} & \textbf{4.69} & \textbf{56.02} & \textbf{4.55} \\
        \midrule

        \multirow{2}{*}{Llama3.2-3B} 
        & random & 87.89   & 86.47   & 76.35 & 68.69 & 90.54 & 90.00 & 75.82 & 40.00 & 94.98 & {69.65} & 95.66 & 85.96 & 92.75 & 76.56 & 93.30 & 81.82 \\
        & \textbf{cognitive} & \textbf{49.47} & \textbf{17.06} & \textbf{49.69} & \textbf{13.13} & \textbf{52.29} & \textbf{10.00} & \textbf{43.62} & \textbf{0.00} & \textbf{92.01} & {80.10} & \textbf{53.60} & \textbf{7.02} & \textbf{46.69} & \textbf{0.00} & \textbf{49.25} & \textbf{0.00} \\
        \midrule

        \multirow{2}{*}{Qwen3-8B} 
        & random & 92.81 & 75.29 & 89.90 & 53.54 & 92.73 & 42.50 & 88.60 & 80.00 & 92.69 & 60.20 & 94.45 & 24.56 & 94.15 & 20.31 & 96.52 & 31.82 \\
        & \textbf{cognitive} & \textbf{59.19} & \textbf{38.24} & \textbf{64.81} & \textbf{30.30} & \textbf{85.95} & {47.50} & \textbf{46.26} & \textbf{0.00} & \textbf{89.29} & \textbf{53.23} & \textbf{72.77} & {35.09} & \textbf{87.61} & {21.88} & \textbf{83.17} & {54.55} \\
        \midrule

        \multirow{2}{*}{Qwen3-4B} 
        & random & 94.17 & 84.71 & 84.61 & 77.78 & 86.91 & 77.50 & 98.15 & 80.00 & 87.15 & 44.78 & 96.89 & 87.72 & 92.00 & 75.00 & 94.79 & 72.73 \\
        & \textbf{cognitive} & \textbf{80.13} & \textbf{64.71} & \textbf{63.10} & \textbf{35.35} & \textbf{65.95} & \textbf{60.00} & \textbf{46.25} & \textbf{0.00} & \textbf{82.40} & {46.27} & \textbf{84.88} & \textbf{64.91} & \textbf{82.79} & \textbf{39.06} & \textbf{45.49} & \textbf{13.64} \\
        \midrule

        \multirow{2}{*}{Yi-1.5-9B}
        & random & 86.83 & 79.41 & 82.02 & 54.55 & 77.40 & 35.00 & 81.53 & 60.00 & 76.04 & 36.32 & 89.83 & 36.84 & 87.53 & 42.19 & 86.27 & 63.64 \\
        & \textbf{cognitive} & \textbf{52.76} & \textbf{21.76} & \textbf{45.99} & \textbf{9.09} & \textbf{47.25} & \textbf{2.50} & \textbf{48.10} & \textbf{40.00} & \textbf{54.22} & \textbf{16.92} & \textbf{52.41} & \textbf{15.79} & \textbf{82.75} & \textbf{26.56} & \textbf{62.85} & \textbf{18.18} \\
        \midrule

        \multirow{2}{*}{Yi-1.5-6B} 
        & random & 80.64 & 69.41 & 68.82 & 38.38 & 77.83 & 55.00 & 69.61 & 60.00 & 73.33 & 43.78 & 77.71 & 22.81 & 81.65 & 29.69 & 88.54 & 72.73 \\
        & \textbf{cognitive} & \textbf{49.90} & \textbf{15.29} & \textbf{68.23} & {41.41} & \textbf{49.54} & \textbf{2.50} & \textbf{42.92} & \textbf{0.00} & {76.64} & {43.78} & \textbf{68.53} & \textbf{14.04} & \textbf{44.94} & \textbf{0.00} & \textbf{86.28} & \textbf{50.00} \\
        \bottomrule
    \end{tabular}
    }
    \label{tab:llm_results}
\end{table*}


\begin{figure*}[]
	\centering
    \includegraphics[width=1\linewidth]{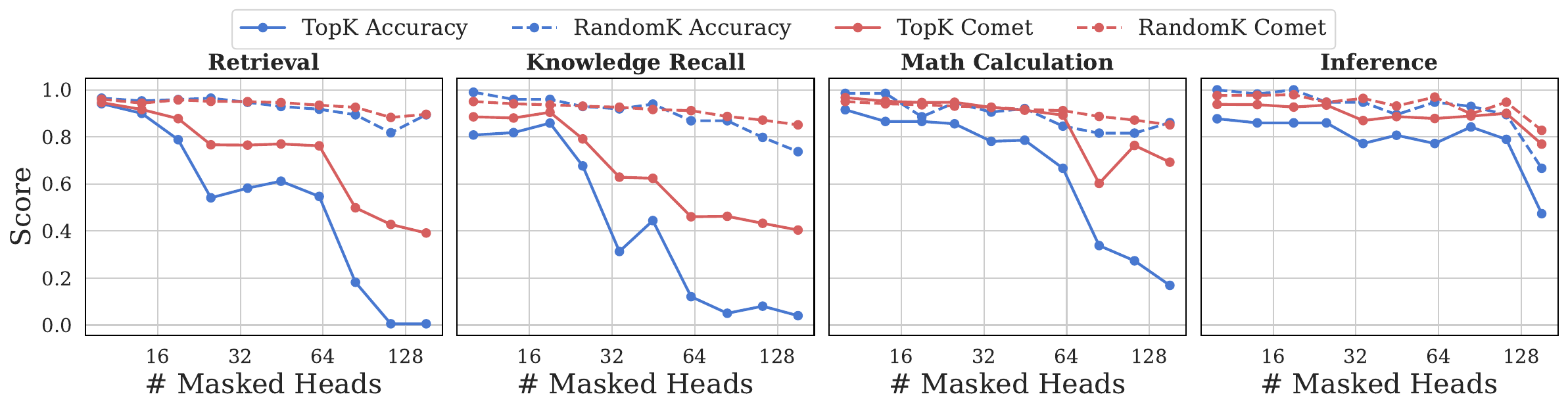}
	\caption{The performance of Llama3.1-8B-instruct by masking out top K cognitive heads vs
K random heads on retrieval, knowledge recall, math calculation, and inference.
	}
	\label{masking_num}
\end{figure*}




\subsection{Functional Contributions of Cognitive Heads}

After identifying the cognitive heads associated with each function, we examine their functional roles by evaluating the model's behavior on the CogQA test set under targeted interventions.
We perform head ablation by scaling the output of a specific attention head with a small factor $\epsilon$ (e.g., 0.001), effectively suppressing its contribution:

\begin{equation}
x_i^{\text{mask}} = \operatorname{Softmax}\left(\frac{W_q^i W_k^{i T}}{\sqrt{d_k / n}}\right) \cdot \epsilon W_v^i
\end{equation}
Specifically, we compare model performance when masking identified cognitive heads versus masking an equal number of randomly selected heads. To quantify the impact of masking, we use several standard evaluation metrics including COMET~\cite{rei2020comet}, BLEU~\cite{papineni2002bleu}, ROUGE~\cite{chin2004rouge}, and semantic similarity to compare the model's outputs before and after intervention.  We define an output as unaffected if the BLEU score exceeds 0.8, or either the ROUGE or semantic similarity scores surpass 0.6, and compute accuracy accordingly.

As shown in Table~\ref{tab:llm_results}, masking cognitive heads leads to a significant decline in performance, whereas masking an equal number of random heads results in only marginal degradation across all LLMs. In some cases, masking the identified cognitive heads causes the accuracy to drop to zero, indicating that the model cannot execute the corresponding function without them. This sharp contrast highlights the essential role cognitive heads play in enabling specific reasoning capabilities. 
\textcolor{black}{To further validate the functional specialization, we conduct experiments where we mask the retrieval heads during the evaluation of knowledge recall (Recall), and conversely, mask knowledge recall heads during the evaluation of retrieval performance. The results in Table \ref{tab:other} show that masking the corresponding cognitive heads causes a significantly larger performance drop than masking others.}

\begin{table}[]
\centering
\caption{Intervention results (\%) of different cognitive heads and random heads across Retrieval and Knowledge Recall functions.}
\resizebox{0.95\linewidth}{!}{
\begin{tabular}{lccccc}
\toprule
 \textbf{Model} & \textbf{Inter\_Head} & \textbf{Retrieval (comet)} & \textbf{Retrieval (acc) } & \textbf{Recall (comet)} & \textbf{Recall (acc)} \\
\midrule
Llama3.1-8B & random & 90.83 & 84.71 & 87.85 & 83.84 \\
Llama3.1-8B & retrieval & \textbf{44.96} & \textbf{8.24} & 72.05 & \textbf{33.33}\\
Llama3.1-8B & recall & 86.79 & 75.29 & \textbf{56.93} & 38.38 \\
\hline
Qwen3-8B & random & 92.81 & 75.29 & 89.90 & 53.54\\
Qwen3-8B & retrieval & \textbf{59.19} & \textbf{38.24} & 79.26 & 57.58 \\
Qwen3-8B & recall & 83.31 & 71.18 & \textbf{64.81} & \textbf{30.30}\\
\bottomrule
\end{tabular}
}
\label{tab:other}
\end{table}

We further investigate the performance of model under different numbers of masked attention heads. As shown in Figure~\ref{masking_num}, increasing the number of randomly masked heads has minimal impact on overall performance of Llama3.1-8B-instruct. In contrast, masking cognitive heads results in a significant drop in performance across various functions. Notably, masking heads associated with Retrieval and Knowledge Recall causes a pronounced degradation in their respective functions, whereas functions such as Math Calculation and Inference exhibit more resilience. This suggests that certain cognitive functions depend more heavily on specific, distinguishable attention heads, while others are distributed more broadly across the model.

\subsection{Relationship Among Cognitive Heads}

While cognitive heads are specialized for distinct functions, understanding their relationships is crucial for revealing how complex reasoning emerges from their cooperation.

\begin{figure}[t]
    \centering
    \begin{subfigure}[t]{0.3\linewidth}
        \centering
        \includegraphics[width=\linewidth]{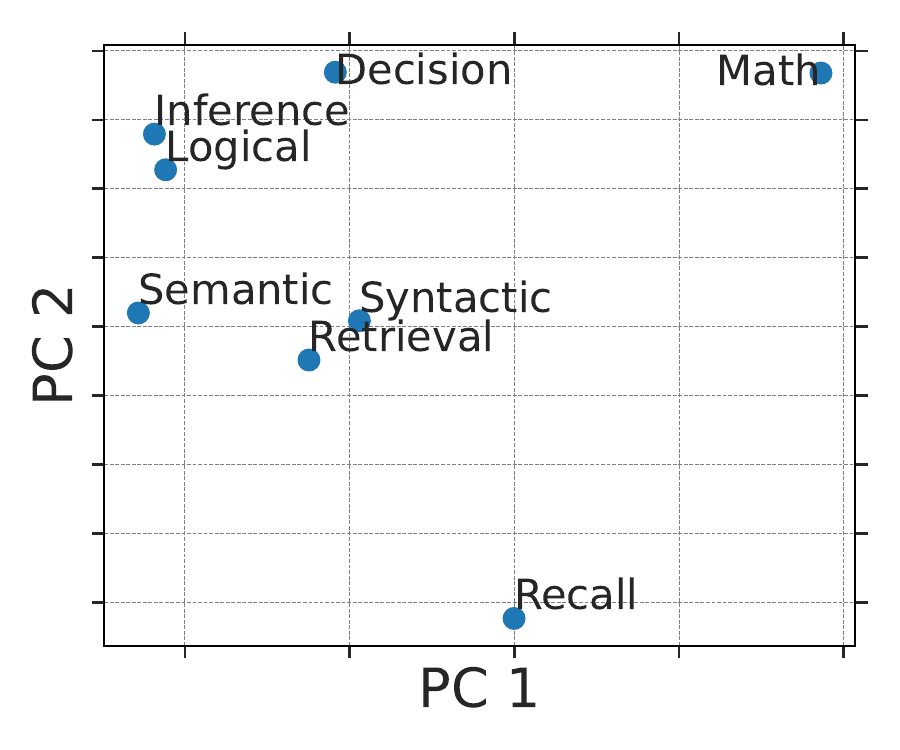}
        \caption{Llama3.1-8B}
        \label{fig:pca-a}
    \end{subfigure}
    \begin{subfigure}[t]{0.3\linewidth}
        \centering
        \includegraphics[width=\linewidth]{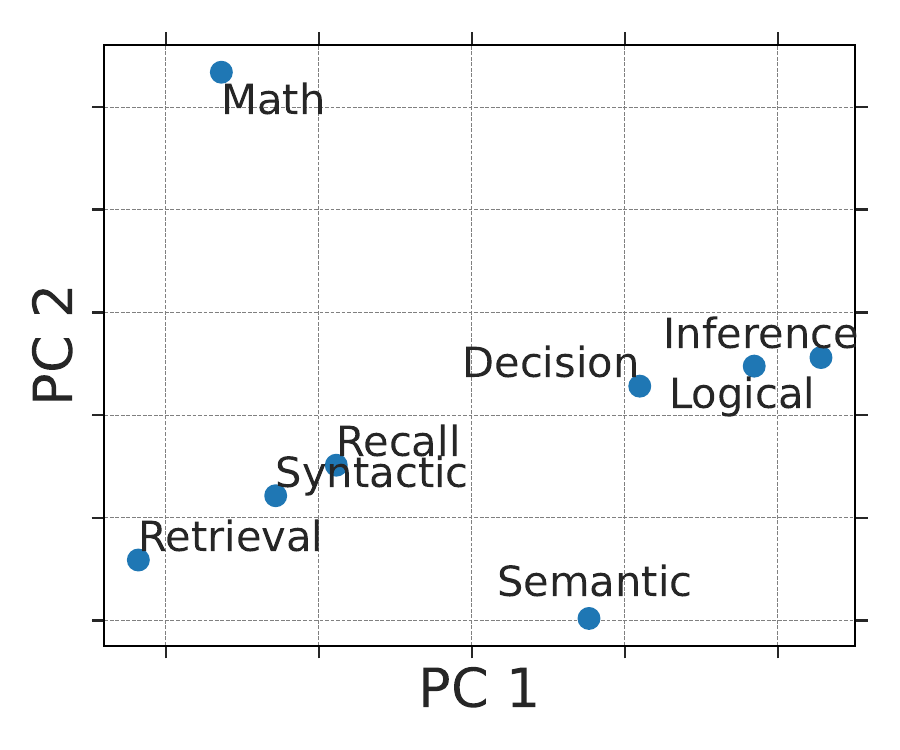}
        \caption{Qwen3-4B}
        \label{fig:pca-b}
    \end{subfigure}
    \begin{subfigure}[t]{0.3\linewidth}
        \centering
        \includegraphics[width=\linewidth]{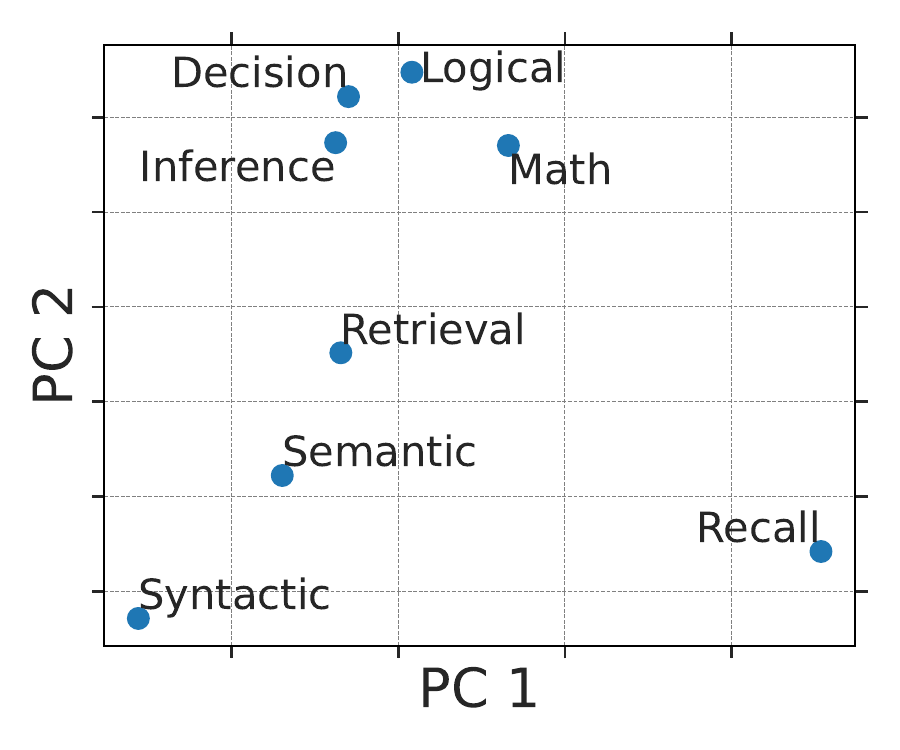}
        \caption{Yi-1.5-6B}
        \label{fig:pca-c}
    \end{subfigure}
    \vspace{-2mm}
    \caption{PCA visualization of the 8 function heads' clustering in three models.}
    \label{fig:pca}
    \vspace{-3mm}
\end{figure}

\textbf{Functional Clustering:} Inspired by neuroscience findings that related cognitive functions localize in overlapping brain regions (e.g., prefrontal cortex for reasoning and inference~\cite{barsalou2014cognitive}), we investigate whether LLM attention heads show similar patterns. We rank each head’s importance across eight cognitive functions, form ranking vectors, and apply principal component analysis (PCA) to visualize their organization (Figure~\ref{fig:pca}). The results reveal clear clustering: heads linked to reasoning, inference, and decision-making group closely, while those related to mathematical computation form a distinct cluster in Llama and Qwen, and lie adjacent to reasoning heads in Yi. Lower-level functions also show moderate clustering. These patterns suggest a modular functional architecture in LLMs akin to that in the human brain.


\definecolor{iceblue}{HTML}{E0F5FF}

\begin{table*}
  \caption{Study on the influence of low-level cognitive heads for high-order function on Llama3.1-8B-instruct. Accuracy is measured based on BLEU, ROUGE, and semantic similarity scores.}
  \centering
  \setlength{\tabcolsep}{2pt}
  \renewcommand{\arraystretch}{1.1}
  \resizebox{0.95\textwidth}{!}{%
    \begin{tabular}{ccccccccc}
    \toprule
     Retrieval & Knowledge & Semantic & Syntactic & Math & Inference & Logical & Decision \\
    \midrule
    \xmark & \cmark & \cmark & \cmark &  
    $0.00_{\tiny \begin{tcolorbox}[colback=iceblue, colframe=iceblue, width=1cm, height=0.35cm, boxrule=0pt, left=0pt, right=0pt, top=0pt, bottom=0pt, halign=center, valign=center]{\text{$\downarrow$ 100}} \end{tcolorbox}}$ &
    $0.00_{\tiny \begin{tcolorbox}[colback=iceblue, colframe=iceblue, width=1cm, height=0.35cm, boxrule=0pt, left=0pt, right=0pt, top=0pt, bottom=0pt, halign=center, valign=center]{\text{$\downarrow$ 100}} \end{tcolorbox}}$ &
    $0.00_{\tiny \begin{tcolorbox}[colback=iceblue, colframe=iceblue, width=1cm, height=0.35cm, boxrule=0pt, left=0pt, right=0pt, top=0pt, bottom=0pt, halign=center, valign=center]{\text{$\downarrow$ 100}} \end{tcolorbox}}$ &
    $0.00_{\tiny \begin{tcolorbox}[colback=iceblue, colframe=iceblue, width=1cm, height=0.35cm, boxrule=0pt, left=0pt, right=0pt, top=0pt, bottom=0pt, halign=center, valign=center]{\text{$\downarrow$ 100}} \end{tcolorbox}}$ \\
    
    \cmark & \xmark & \cmark & \cmark &  
    $0.00_{\tiny \begin{tcolorbox}[colback=iceblue, colframe=iceblue, width=1cm, height=0.35cm, boxrule=0pt, left=0pt, right=0pt, top=0pt, bottom=0pt, halign=center, valign=center]{\text{$\downarrow$ 100}} \end{tcolorbox}}$ &
    $0.00_{\tiny \begin{tcolorbox}[colback=iceblue, colframe=iceblue, width=1cm, height=0.35cm, boxrule=0pt, left=0pt, right=0pt, top=0pt, bottom=0pt, halign=center, valign=center]{\text{$\downarrow$ 100}} \end{tcolorbox}}$ &
    $0.00_{\tiny \begin{tcolorbox}[colback=iceblue, colframe=iceblue, width=1cm, height=0.35cm, boxrule=0pt, left=0pt, right=0pt, top=0pt, bottom=0pt, halign=center, valign=center]{\text{$\downarrow$ 100}} \end{tcolorbox}}$ &
    $0.00_{\tiny \begin{tcolorbox}[colback=iceblue, colframe=iceblue, width=1cm, height=0.35cm, boxrule=0pt, left=0pt, right=0pt, top=0pt, bottom=0pt, halign=center, valign=center]{\text{$\downarrow$ 100}} \end{tcolorbox}}$ \\
    
    \cmark & \cmark & \xmark & \cmark &
    $66.67_{\tiny \begin{tcolorbox}[colback=iceblue, colframe=iceblue, width=1cm, height=0.35cm, boxrule=0pt, left=0pt, right=0pt, top=0pt, bottom=0pt, halign=center, valign=center]{\text{$\downarrow$ 33.33}} \end{tcolorbox}}$ &
    $88.24_{\tiny \begin{tcolorbox}[colback=iceblue, colframe=iceblue, width=1cm, height=0.35cm, boxrule=0pt, left=0pt, right=0pt, top=0pt, bottom=0pt, halign=center, valign=center]{\text{$\downarrow$ 11.76}} \end{tcolorbox}}$ &
    $93.10_{\tiny \begin{tcolorbox}[colback=iceblue, colframe=iceblue, width=1cm, height=0.35cm, boxrule=0pt, left=0pt, right=0pt, top=0pt, bottom=0pt, halign=center, valign=center]{\text{$\downarrow$ 8.90}} \end{tcolorbox}}$ &
    $57.14_{\tiny \begin{tcolorbox}[colback=iceblue, colframe=iceblue, width=1cm, height=0.35cm, boxrule=0pt, left=0pt, right=0pt, top=0pt, bottom=0pt, halign=center, valign=center]{\text{$\downarrow$ 42.86}} \end{tcolorbox}}$ \\
    
    \cmark & \cmark & \cmark & \xmark & - &
    $76.92_{\tiny \begin{tcolorbox}[colback=iceblue, colframe=iceblue, width=1cm, height=0.35cm, boxrule=0pt, left=0pt, right=0pt, top=0pt, bottom=0pt, halign=center, valign=center]{\text{$\downarrow$ 23.08}} \end{tcolorbox}}$ &
    $100_{\tiny \begin{tcolorbox}[colback=iceblue, colframe=iceblue, width=1cm, height=0.35cm, boxrule=0pt, left=0pt, right=0pt, top=0pt, bottom=0pt, halign=center, valign=center]{\text{ 0.00}} \end{tcolorbox}}$ &
    $100_{\tiny \begin{tcolorbox}[colback=iceblue, colframe=iceblue, width=1cm, height=0.35cm, boxrule=0pt, left=0pt, right=0pt, top=0pt, bottom=0pt, halign=center, valign=center]{\text{ 0.00}} \end{tcolorbox}}$ \\
    
    \bottomrule
    \end{tabular}
  }
  \label{tab:abl_level}
\end{table*}
\textbf{Hierarchical Structure:} Human problem solving often involves hierarchical reasoning, where lower-level functions such as retrieval and comprehension support higher-level inference and decision-making. The CogQA dataset captures this structure through subquestions progressing from simple information extraction to complex reasoning. We test if LLMs reflect this hierarchy by masking attention heads tied to early-stage functions and measuring the effect on later tasks. \textcolor{black}{For instance, to assess how Retrieval affects Math Calculation, we suppress Retrieval-related heads throughout the subquestions. Answers from earlier Retrieval are used as priors for later math reasoning, allowing us to observe how disrupting low-level functions can propagate and impair higher-level reasoning along the chain.} As Table~\ref{tab:abl_level} shows, masking retrieval or knowledge recall heads causes significant performance drops in subsequent decision-making steps, whereas masking syntactic understanding heads has minimal impact. This provides evidence for an emergent hierarchical organization in LLMs, where foundational cognitive functions underpin advanced reasoning.

\subsection{Influence of Cognitive Heads on Downstream Tasks}

In this section, we investigate how cognitive heads influence downstream tasks through both negative interventions (masking out cognitive function heads) and positive interventions (shifting heads toward specific functions). 
\textcolor{black}{We conduct experiments on two tasks: a math task using 100 GSM8K samples (GSM8K\_100) and a retrieval task with 49 samples from an extractive\_QA dataset. The Extractive\_QA pairs are generated by GPT-4o, with answers extracted directly from the source paragraph.}

\textbf{Negative Intervention:}
\textcolor{black}{We perform negative intervention by masking corresponding cognitive heads (Math Calculation heads for GSM8K\_100 and Retrieval heads for Extractive\_QA), effectively suppressing their activations. As shown in Table \ref{tab:negative}, this causes significant performance drops across models, confirming these heads’ functional roles. Notably, after masking, performance converges to a similarly low level across different LLMs, regardless of model size or original accuracy. This is expected, as the crucial cognitive heads responsible for specific functions are disabled, making it difficult for the model to arrive at correct answers.}

\textcolor{black}{For math, the remaining 30\% accuracy likely stems from two factors: (1) memorized answers in the base model, and (2) simple questions not requiring actual computation.
For retrieval, masking Retrieval heads almost completely abolishes the model’s retrieval ability across all scales. This indicates that cognitive functions are indeed localized in a subset of heads, and masking them leads to a systematic degradation, irrespective of model capacity.} The negative intervention example further shows that, masking the \textbf{Math Calculation} heads leads to errors in arithmetic tasks, while retrieval and language functions remain largely unaffected. This confirms that these cognitive heads are crucial for specific functions and highlights the robustness and generalizability of our method.

\textbf{Positive Intervention:} We calculate the activation directions of different cognitive functions using the CogQA dataset. For each function, the activation direction of a head at layer $l$ and index $h$ is computed as:
\begin{equation}
\operatorname{dir}_l^h=\mathbb{E}_{i \in \mathcal{D}_{\text {correct }}}\left[x_l^h(i)\right]-\mathbb{E}_{i \in \mathcal{D}_{\text {incorrect }}}\left[x_l^h(i)\right]
\end{equation}
where $x_l^h(i)$ denotes the activation of head at layer $l$ and index $h$, and $\mathcal{D}_{\text {correct }}$ and $\mathcal{D}_{\text {incorrect }}$ represent the sets of samples answered correctly and incorrectly, respectively. Then we estimate the standard deviation of activations \cite{truthful} along the cognitive function direction to be $\sigma_l^h$, and shift original head activation as $x_l^h(i) \leftarrow x_l^h(i)+\alpha \sigma_l^h \operatorname{dir}_l^h$, where $\alpha$ is a parameter. 

\textcolor{black}{The experimental results in Table \ref{tab:negative} show that enhancing the activation of retrieval heads along their corresponding functional directions leads to improved performance on the retrieval task. For the math task, questions often engage multiple cognitive functions. We observe cases where computations are correct but semantic understanding fails. By positively intervening on either Math Calculation or Semantic Understanding heads, performance consistently improves, highlighting the collaborative nature of cognitive functions in complex reasoning and the need for joint interventions—a direction we leave for future work.}

The example below illustrates a positive intervention on \textbf{Math Calculation} heads in Qwen3-4B: the model initially produces an incorrect answer (red) but generates the correct one (blue) after intervention. This demonstrates that adjusting cognitive head activations along their functional directions reliably enhances performance on complex reasoning tasks.

\newtcolorbox{promptbox2}[1][]{
  colback=gray!5!white,
  colframe=black!75!white,
  fonttitle=\bfseries,
  title=Negative Intervention Case,
  #1
}

\begin{promptbox2}
\label{example2}
\textbf{Query:} Question: Solve -42*r + 27*c = -1167 and 130*r + 4*c = 372 for r.

\textbf{Original Output:} Okay, so I need to solve this system of equations: -42r + 27c = -1167 and 130r + 4c = 372
...
First equation: -42*4 + 27*(-37) = -168 - 999 = -1167. Correct.
Second equation: 130*4 + 4*(-37) = 520 - 148 = 372. Correct. Yes, that's right. So everything checks out.
So I think that's the correct answer. Therefore, \textcolor{blue}{r is 4}.
\textbf{Negative Intervention (Math) Output:}
...
To eliminate \( c \), I need the coefficients of \( c \) in both equations to be the same (or negatives). The coefficients are 27 and 4. The least common multiple of 27 and 4 is 100? Wait, 27 is \( 3^3 \), and 4 is \( 2^2 \), so LCM is \( 2^2*3^3 = \textcolor{red}{4*27 = 100} \). So if I can make the coefficients of \( c \) to be 100 and 100, or 27 and 4. Wait, but 27 and 4. So if I multiply the first equation by 4 and the second equation by 27, then the coefficients of \( c \) would be \textcolor{red}{27*4 = 100} and \textcolor{red}{4*27 = 100}. Then, if I subtract the equations, the \( c \) terms would be 100 and 100, but since they are both positive, subtracting the equations would eliminate \( c \).
...
\end{promptbox2}

\newtcolorbox{promptbox1}[1][]{
  colback=gray!5!white,
  colframe=black!75!white,
  fonttitle=\bfseries,
  title=Positive Intervention Case,
  #1
}
\begin{promptbox1}\label{example1}
\textbf{Query:} On a school trip to the seashore, Alan and his friends collected shells. Alan collected four times as many shells as Ben did. Ben got a late start and only collected a third of what Laurie did. If Laurie collected 36 shells how many did Alan collect?

\textbf{Original Output:} Alan collected four times as many shells as Ben did, and Ben collected a third of what Laurie did. Since Laurie collected 36 shells, Ben collected 36 / 3 = 12 shells. Therefore, Alan collected 4 * 12 = 48 shells. However, the question asks for the number of shells Alan collected, and it is given that Alan collected four times as many shells as Ben did. Therefore, the correct answer is 4 * 12 = 48 * 4 / 3 = \textcolor{red}{72}.

\textbf{Positive Intervention (Math) Output :}
...Since Ben collected a third of what Laurie did, he collected 36 / 3 = 12 shells. Alan collected four times as many shells as Ben did, so he collected \textcolor{blue}{4 * 12 = 48 shells}...The correct answer is \textcolor{blue}{48}
\end{promptbox1}

\begin{table}[]
\vspace{-0.3cm}
\centering
\caption{The LLM performance on GSM8k\_{100} and Extractive\_QA by Negative and Positive Intervention (Inter.). Math accuracy (\%) is measured by exact answer match; Extractive\_QA accuracy checks if the original paragraph answer appears in the response.}
\resizebox{0.9\linewidth}{!}{
\begin{tabular}{lcccccc}
\toprule
 \textbf{Dataset} & \textbf{Method} & \textbf{Head} & \textbf{Llama3.1-8B} & \textbf{Llama3.2-3B} & \textbf{Qwen3-8B} & \textbf{Qwen3-4B} \\
\midrule
\multirow{3}{*}{Extractive\_QA} & 
 Base & - & 57.14 & 36.73 & 57.14 & 51.02\\
& Negative Inter. & Retrieval & 0 & 0 & 14.29 & 12.24\\
& Positive Inter. & Retrieval & 63.26 & 44.90 & 61.22 & 69.38  \\
 \midrule
\multirow{4}{*}{GSM8K\_{100}}  
& Base & - & 82 & 64 & 94 & 91 \\
& Negative Inter. & Math & 38 & 34 & 34 & 37 \\
& Positive Inter. & Math & 84 & 66 & 94 & 92 \\
& Positive Inter. & Semantic & 84 & 65 & 94 & 93 \\
\bottomrule
\end{tabular}
}
\label{tab:negative}
\vspace{-0.3cm}
\end{table}

\section{Related Works}

\paragraph{Neural Networks and the Brain} Neural networks have long been studied as computational models of the brain, with early work linking artificial neurons to the biological mechanisms of perception and learning~\cite{McCullochP90}. Convolutional neural networks (CNNs), in particular, have been shown to capture representations similar to those in the visual cortex~\cite{yamins2014performance}, and more recent studies suggest that the functional modularity observed in deep networks gives rise to brain-like specialization~\citep{dobs2022brain} in vision task. More recently, LLMs have exhibited striking parallels with human brain activity during language processing. In particular, transformer-based models, such as GPT-2, produce internal representations that align with neural responses in language-selective brain regions~\cite{caucheteux2022deep, schrimpf2021neural}.  
However, prior work mostly focuses on perception and language representations, with limited study on higher-level cognitive functions like reasoning. We instead analyze LLMs’ behavior in complex reasoning tasks to explore their alignment with human cognitive functions and functional specialization.

\paragraph{Functional Specialization of Attention Heads}
Recent years have witnessed growing interest in understanding the functional roles of attention heads in Transformer-based models, forming a core component of mechanistic interpretability research. Early work by~\cite{clark2019does} demonstrated that individual heads in BERT capture specific linguistic phenomena such as syntactic dependencies and coreference, indicating a degree of functional specialization. Building on this, \cite{voita2019analyzing} proposed a pruning-based approach to identify important heads by measuring their contribution to downstream performance, showing that many heads are redundant. Subsequent studies extended this analysis to decoder-only large language models (LLMs). \cite{michel2019sixteen} explored functional decomposition in such models, leading to the identification of distinct attention heads responsible for tasks such as pattern induction~\citep{induction}, truthfulness \citep{truthful}, information retrieval \citep{wu2404retrieval}, and safety alignment \citep{safety}. For a broader survey, see \cite{zheng2409attention}.
Despite these advances, most prior work focuses on isolated heads and evaluates them in relatively simple or synthetic tasks. In contrast, we investigate functionally specialized heads under more complex reasoning settings by aligning attention head behavior with human cognitive functions.

\section{Limitations and Future works}

While our study provides an initial framework for analyzing the cognitive functions of attention heads, several limitations remain. First, we focus on eight predefined cognitive functions, which, though representative, may not capture the full spectrum of LLM capabilities; future work could extend this taxonomy with finer-grained or emergent functions. Each subquestion in CogVision is annotated with a single cognitive function, though real reasoning may engage multiple functions. Similarly, we assume one head corresponds to one function, while in practice a head may support multiple functions, vary with context, or reflect hierarchical compositions. These complexities are not fully addressed in our current framework. Excluding subquestions with incorrect subanswers could improve multi-class probing, and further investigation is needed to understand heads serving multiple functions. Finally, our work emphasizes analysis over application, but identifying cognitively relevant heads could inform model design, including dynamic head activation, improved chain-of-thought prompting, targeted fine-tuning, or modular architectures—directions we leave for future exploration.

\section{Conclusions}
We propose an interpretability framework that connects attention heads in large language models (LLMs) to human cognitive functions involved in reasoning. To support this, we introduce CogQA, a cognitively grounded dataset, along with a multi-class classification approach to identify specialized heads associated with specific reasoning tasks. Our analysis across multiple LLM families and scales demonstrates that attention heads exhibit universality, sparsity, intrinsic roles, and dynamic, hierarchical organization. These findings indicate that LLMs internally organize reasoning processes in a manner akin to human cognition, laying the groundwork for more interpretable and cognitively informed language models.











\section*{Acknowledgements}
This work is partially supported by the following Australian Research Council (ARC) projects: FT220100318, DP220102121, LP220100527, LP220200949, DP230101534.

\bibliography{sample_base}

@String{Computing = "Computing" }

@String{Computer = "{IEEE} Computer" }

@article{wu2404retrieval,
  title={Retrieval Head Mechanistically Explains Long-Context Factuality, 2024},
  author={Wu, Wenhao and Wang, Yizhong and Xiao, Guangxuan and Peng, Hao and Fu, Yao},
  journal={URL https://arxiv. org/abs/2404},
  volume={15574}
}

@article{truthful,
  title={Inference-time intervention: Eliciting truthful answers from a language model},
  author={Li, Kenneth and Patel, Oam and Vi{\'e}gas, Fernanda and Pfister, Hanspeter and Wattenberg, Martin},
  journal={Advances in Neural Information Processing Systems},
  volume={36},
  pages={41451--41530},
  year={2023}
}

@article{alain2016understanding,
  title={Understanding intermediate layers using linear classifier probes},
  author={Alain, Guillaume and Bengio, Yoshua},
  journal={arXiv preprint arXiv:1610.01644},
  year={2016}
}

@article{belinkov2022probing,
  title={Probing classifiers: Promises, shortcomings, and advances},
  author={Belinkov, Yonatan},
  journal={Computational Linguistics},
  volume={48},
  number={1},
  pages={207--219},
  year={2022},
  publisher={MIT Press One Broadway, 12th Floor, Cambridge, Massachusetts 02142, USA~…}
}

@article{tenney2019bert,
  title={BERT rediscovers the classical NLP pipeline},
  author={Tenney, Ian and Das, Dipanjan and Pavlick, Ellie},
  journal={arXiv preprint arXiv:1905.05950},
  year={2019}
}

@article{zheng2409attention,
  title={Attention heads of large language models: A survey. arXiv 2024},
  author={Zheng, Z and Wang, Y and Huang, Y and Song, S and Yang, M and Tang, B and Xiong, F and Li, Z},
  journal={arXiv preprint arXiv:2409.03752}
}

@article{achiam2023gpt,
  title={Gpt-4 technical report},
  author={Achiam, Josh and Adler, Steven and Agarwal, Sandhini and Ahmad, Lama and Akkaya, Ilge and Aleman, Florencia Leoni and Almeida, Diogo and Altenschmidt, Janko and Altman, Sam and Anadkat, Shyamal and others},
  journal={arXiv preprint arXiv:2303.08774},
  year={2023}
}

@article{touvron2023llama,
  title={Llama: Open and efficient foundation language models},
  author={Touvron, Hugo and Lavril, Thibaut and Izacard, Gautier and Martinet, Xavier and Lachaux, Marie-Anne and Lacroix, Timoth{\'e}e and Rozi{\`e}re, Baptiste and Goyal, Naman and Hambro, Eric and Azhar, Faisal and others},
  journal={arXiv preprint arXiv:2302.13971},
  year={2023}
}

@article{grattafiori2024llama,
  title={The llama 3 herd of models},
  author={Grattafiori, Aaron and Dubey, Abhimanyu and Jauhri, Abhinav and Pandey, Abhinav and Kadian, Abhishek and Al-Dahle, Ahmad and Letman, Aiesha and Mathur, Akhil and Schelten, Alan and Vaughan, Alex and others},
  journal={arXiv preprint arXiv:2407.21783},
  year={2024}
}

@article{yang2024qwen2,
  title={Qwen2. 5 technical report},
  author={Yang, An and Yang, Baosong and Zhang, Beichen and Hui, Binyuan and Zheng, Bo and Yu, Bowen and Li, Chengyuan and Liu, Dayiheng and Huang, Fei and Wei, Haoran and others},
  journal={arXiv preprint arXiv:2412.15115},
  year={2024}
}

@article{induction,
  title={In-context learning and induction heads},
  author={Olsson, Catherine and Elhage, Nelson and Nanda, Neel and Joseph, Nicholas and DasSarma, Nova and Henighan, Tom and Mann, Ben and Askell, Amanda and Bai, Yuntao and Chen, Anna and others},
  journal={arXiv preprint arXiv:2209.11895},
  year={2022}
}

@article{safety,
  title={On the Role of Attention Heads in Large Language Model Safety},
  author={Zhou, Zhenhong and Yu, Haiyang and Zhang, Xinghua and Xu, Rongwu and Huang, Fei and Wang, Kun and Liu, Yang and Fang, Junfeng and Li, Yongbin},
  journal={arXiv preprint arXiv:2410.13708},
  year={2024}
}

@book{barsalou2014cognitive,
  title={Cognitive psychology: An overview for cognitive scientists},
  author={Barsalou, Lawrence W},
  year={2014},
  publisher={Psychology Press}
}

@article{ono2022bidirectional,
  title={Bidirectional connectivity between Broca's area and Wernicke's area during interactive verbal communication},
  author={Ono, Yumie and Zhang, Xian and Noah, J Adam and Dravida, Swethasri and Hirsch, Joy},
  journal={Brain connectivity},
  volume={12},
  number={3},
  pages={210--222},
  year={2022},
  publisher={Mary Ann Liebert, Inc., publishers 140 Huguenot Street, 3rd Floor New~…}
}

@article{vaswani2017attention,
  title={Attention is all you need},
  author={Vaswani, Ashish and Shazeer, Noam and Parmar, Niki and Uszkoreit, Jakob and Jones, Llion and Gomez, Aidan N and Kaiser, {\L}ukasz and Polosukhin, Illia},
  journal={Advances in neural information processing systems},
  volume={30},
  year={2017}
}

@article{diamond2013executive,
  title={Executive functions},
  author={Diamond, Adele},
  journal={Annual review of psychology},
  volume={64},
  number={1},
  pages={135--168},
  year={2013},
  publisher={Annual Reviews}
}

@book{anderson2014rules,
  title={Rules of the mind},
  author={Anderson, John R},
  year={2014},
  publisher={Psychology Press}
}

@article{hurst2024gpt,
  title={Gpt-4o system card},
  author={Hurst, Aaron and Lerer, Adam and Goucher, Adam P and Perelman, Adam and Ramesh, Aditya and Clark, Aidan and Ostrow, AJ and Welihinda, Akila and Hayes, Alan and Radford, Alec and others},
  journal={arXiv preprint arXiv:2410.21276},
  year={2024}
}

@misc{o4mini2024,
  author       = {OpenAI},
  title        = {Introducing O3 and O4 Mini},
  year         = {2024},
  howpublished = {\url{https://openai.com/index/introducing-o3-and-o4-mini/}},
  note         = {Accessed: 2025-05-05}
}

@article{wheeler1997toward,
  title={Toward a theory of episodic memory: the frontal lobes and autonoetic consciousness.},
  author={Wheeler, Mark A and Stuss, Donald T and Tulving, Endel},
  journal={Psychological bulletin},
  volume={121},
  number={3},
  pages={331},
  year={1997},
  publisher={American Psychological Association}
}

@article{meyer2005language,
  title={Language processing within the human medial temporal lobe},
  author={Meyer, Patric and Mecklinger, Axel and Grunwald, Thomas and Fell, Juergen and Elger, Christian E and Friederici, Angela D},
  journal={Hippocampus},
  volume={15},
  number={4},
  pages={451--459},
  year={2005},
  publisher={Wiley Online Library}
}

@article{hubbard2005interactions,
  title={Interactions between number and space in parietal cortex},
  author={Hubbard, Edward M and Piazza, Manuela and Pinel, Philippe and Dehaene, Stanislas},
  journal={Nature reviews neuroscience},
  volume={6},
  number={6},
  pages={435--448},
  year={2005},
  publisher={Nature Publishing Group UK London}
}

@article{young2024yi,
  title={Yi: Open foundation models by 01. ai},
  author={Young, Alex and Chen, Bei and Li, Chao and Huang, Chengen and Zhang, Ge and Zhang, Guanwei and Wang, Guoyin and Li, Heng and Zhu, Jiangcheng and Chen, Jianqun and others},
  journal={arXiv preprint arXiv:2403.04652},
  year={2024}
}

@inproceedings{creak,
  author       = {Yasumasa Onoe and
                  Michael J. Q. Zhang and
                  Eunsol Choi and
                  Greg Durrett},
  title        = {{CREAK:} {A} Dataset for Commonsense Reasoning over Entity Knowledge},
  booktitle    = {Proceedings of the Neural Information Processing Systems Track on
                  Datasets and Benchmarks 1, NeurIPS Datasets and Benchmarks 2021, December
                  2021, virtual},
  year         = {2021},
  url          = {https://datasets-benchmarks-proceedings.neurips.cc/paper/2021/hash/5737c6ec2e0716f3d8a7a5c4e0de0d9a-Abstract-round2.html},
  bibsource    = {dblp computer science bibliography, https://dblp.org}
}

@inproceedings{ecqa,
  title={{E}xplanations for {C}ommonsense{QA}: {N}ew {D}ataset and {M}odels},
  author={Shourya Aggarwal and Divyanshu Mandowara and Vishwajeet Agrawal and Dinesh Khandelwal and Parag Singla and Dinesh Garg},
  booktitle="Proceedings of the 59th Annual Meeting of the Association for Computational Linguistics and the 11th International Joint Conference on Natural Language Processing (Volume 1: Long Papers)",
  year = "2021",
  address = "Online",
  publisher = "Association for Computational Linguistics"
}

@inproceedings{esnli,
  author       = {Oana{-}Maria Camburu and
                  Tim Rockt{\"{a}}schel and
                  Thomas Lukasiewicz and
                  Phil Blunsom},
  editor       = {Samy Bengio and
                  Hanna M. Wallach and
                  Hugo Larochelle and
                  Kristen Grauman and
                  Nicol{\`{o}} Cesa{-}Bianchi and
                  Roman Garnett},
  title        = {e-SNLI: Natural Language Inference with Natural Language Explanations},
  booktitle    = {Advances in Neural Information Processing Systems 31: Annual Conference
                  on Neural Information Processing Systems 2018, NeurIPS 2018, December
                  3-8, 2018, Montr{\'{e}}al, Canada},
  pages        = {9560--9572},
  year         = {2018},
  url          = {https://proceedings.neurips.cc/paper/2018/hash/4c7a167bb329bd92580a99ce422d6fa6-Abstract.html},
  bibsource    = {dblp computer science bibliography, https://dblp.org}
}

@article{gsm8k,
  author       = {Karl Cobbe and
                  Vineet Kosaraju and
                  Mohammad Bavarian and
                  Mark Chen and
                  Heewoo Jun and
                  Lukasz Kaiser and
                  Matthias Plappert and
                  Jerry Tworek and
                  Jacob Hilton and
                  Reiichiro Nakano and
                  Christopher Hesse and
                  John Schulman},
  title        = {Training Verifiers to Solve Math Word Problems},
  journal      = {CoRR},
  volume       = {abs/2110.14168},
  year         = {2021},
  url          = {https://arxiv.org/abs/2110.14168},
  eprinttype    = {arXiv},
  eprint       = {2110.14168},
  bibsource    = {dblp computer science bibliography, https://dblp.org}
}

@inproceedings{aqua,
  title={Program Induction by Rationale Generation: Learning to Solve and Explain Algebraic Word Problems},
  author={Ling, Wang and Yogatama, Dani and Dyer, Chris and Blunsom, Phil},
  booktitle={ACL},
  year={2017}
}

@inproceedings{llm_annotate,
  author       = {Xinru Wang and
                  Hannah Kim and
                  Sajjadur Rahman and
                  Kushan Mitra and
                  Zhengjie Miao},
  editor       = {Florian 'Floyd' Mueller and
                  Penny Kyburz and
                  Julie R. Williamson and
                  Corina Sas and
                  Max L. Wilson and
                  Phoebe O. Toups Dugas and
                  Irina Shklovski},
  title        = {Human-LLM Collaborative Annotation Through Effective Verification
                  of {LLM} Labels},
  booktitle    = {Proceedings of the {CHI} Conference on Human Factors in Computing
                  Systems, {CHI} 2024, Honolulu, HI, USA, May 11-16, 2024},
  pages        = {303:1--303:21},
  publisher    = {{ACM}},
  year         = {2024},
  url          = {https://doi.org/10.1145/3613904.3641960},
  doi          = {10.1145/3613904.3641960},
  bibsource    = {dblp computer science bibliography, https://dblp.org}
}

@inproceedings{papineni2002bleu,
  title={Bleu: a method for automatic evaluation of machine translation},
  author={Papineni, Kishore and Roukos, Salim and Ward, Todd and Zhu, Wei-Jing},
  booktitle={Proceedings of the 40th annual meeting of the Association for Computational Linguistics},
  pages={311--318},
  year={2002}
}

@inproceedings{chin2004rouge,
  title={Rouge: A package for automatic evaluation of summaries},
  author={Chin-Yew, Lin},
  booktitle={Proceedings of the Workshop on Text Summarization Branches Out, 2004},
  year={2004}
}

@article{rei2020comet,
  title={COMET: A neural framework for MT evaluation},
  author={Rei, Ricardo and Stewart, Craig and Farinha, Ana C and Lavie, Alon},
  journal={arXiv preprint arXiv:2009.09025},
  year={2020}
}

@article{clark2019does,
  title={What does bert look at? an analysis of bert's attention},
  author={Clark, Kevin and Khandelwal, Urvashi and Levy, Omer and Manning, Christopher D},
  journal={arXiv preprint arXiv:1906.04341},
  year={2019}
}

@article{voita2019analyzing,
  title={Analyzing multi-head self-attention: Specialized heads do the heavy lifting, the rest can be pruned},
  author={Voita, Elena and Talbot, David and Moiseev, Fedor and Sennrich, Rico and Titov, Ivan},
  journal={arXiv preprint arXiv:1905.09418},
  year={2019}
}

@article{michel2019sixteen,
  title={Are sixteen heads really better than one?},
  author={Michel, Paul and Levy, Omer and Neubig, Graham},
  journal={Advances in neural information processing systems},
  volume={32},
  year={2019}
}

@inproceedings{cot,
  author       = {Jason Wei and
                  Xuezhi Wang and
                  Dale Schuurmans and
                  Maarten Bosma and
                  Brian Ichter and
                  Fei Xia and
                  Ed H. Chi and
                  Quoc V. Le and
                  Denny Zhou},
  editor       = {Sanmi Koyejo and
                  S. Mohamed and
                  A. Agarwal and
                  Danielle Belgrave and
                  K. Cho and
                  A. Oh},
  title        = {Chain-of-Thought Prompting Elicits Reasoning in Large Language Models},
  booktitle    = {Advances in Neural Information Processing Systems 35: Annual Conference
                  on Neural Information Processing Systems 2022, NeurIPS 2022, New Orleans,
                  LA, USA, November 28 - December 9, 2022},
  year         = {2022},
  url          = {http://papers.nips.cc/paper\_files/paper/2022/hash/9d5609613524ecf4f15af0f7b31abca4-Abstract-Conference.html},
  bibsource    = {dblp computer science bibliography, https://dblp.org}
}

@incollection{McCullochP90,
  author       = {Warren S. McCulloch and
                  Walter H. Pitts},
  editor       = {Margaret A. Boden},
  title        = {A Logical Calculus of the Ideas Immanent in Nervous Activity},
  booktitle    = {The Philosophy of Artificial Intelligence},
  series       = {Oxford readings in philosophy},
  pages        = {22--39},
  publisher    = {Oxford University Press},
  year         = {1990},
  bibsource    = {dblp computer science bibliography, https://dblp.org}
}

@article{yamins2014performance,
  title={Performance-optimized hierarchical models predict neural responses in higher visual cortex},
  author={Yamins, Daniel LK and Hong, Ha and Cadieu, Charles F and Solomon, Ethan A and Seibert, Darren and DiCarlo, James J},
  journal={Proceedings of the national academy of sciences},
  volume={111},
  number={23},
  pages={8619--8624},
  year={2014},
  publisher={National Academy of Sciences}
}

@article{caucheteux2022deep,
  title={Deep language algorithms predict semantic comprehension from brain activity},
  author={Caucheteux, Charlotte and Gramfort, Alexandre and King, Jean-R{\'e}mi},
  journal={Scientific reports},
  volume={12},
  number={1},
  pages={16327},
  year={2022},
  publisher={Nature Publishing Group UK London}
}

@article{schrimpf2021neural,
  title={The neural architecture of language: Integrative modeling converges on predictive processing},
  author={Schrimpf, Martin and Blank, Idan Asher and Tuckute, Greta and Kauf, Carina and Hosseini, Eghbal A and Kanwisher, Nancy and Tenenbaum, Joshua B and Fedorenko, Evelina},
  journal={Proceedings of the National Academy of Sciences},
  volume={118},
  number={45},
  pages={e2105646118},
  year={2021},
  publisher={National Academy of Sciences}
}

@article{dobs2022brain,
  title={Brain-like functional specialization emerges spontaneously in deep neural networks},
  author={Dobs, Katharina and Martinez, Julio and Kell, Alexander JE and Kanwisher, Nancy},
  journal={Science advances},
  volume={8},
  number={11},
  pages={eabl8913},
  year={2022},
  publisher={American Association for the Advancement of Science}
}
\bibliographystyle{plain}

\section*{NeurIPS Paper Checklist}

\begin{enumerate}

\item {\bf Claims}
    \item[] Question: Do the main claims made in the abstract and introduction accurately reflect the paper's contributions and scope?
    \item[] Answer: \answerYes{} 
    \item[] Justification: Yes, the main claims in the abstract and introduction accurately reflect the paper’s contributions and scope. 
    \item[] Guidelines:
    \begin{itemize}
        \item The answer NA means that the abstract and introduction do not include the claims made in the paper.
        \item The abstract and/or introduction should clearly state the claims made, including the contributions made in the paper and important assumptions and limitations. A No or NA answer to this question will not be perceived well by the reviewers. 
        \item The claims made should match theoretical and experimental results, and reflect how much the results can be expected to generalize to other settings. 
        \item It is fine to include aspirational goals as motivation as long as it is clear that these goals are not attained by the paper. 
    \end{itemize}

\item {\bf Limitations}
    \item[] Question: Does the paper discuss the limitations of the work performed by the authors?
    \item[] Answer: \answerYes{} 
    \item[] Justification: We discussed our limitation in discussion section.
    \item[] Guidelines:
    \begin{itemize}
        \item The answer NA means that the paper has no limitation while the answer No means that the paper has limitations, but those are not discussed in the paper. 
        \item The authors are encouraged to create a separate "Limitations" section in their paper.
        \item The paper should point out any strong assumptions and how robust the results are to violations of these assumptions (e.g., independence assumptions, noiseless settings, model well-specification, asymptotic approximations only holding locally). The authors should reflect on how these assumptions might be violated in practice and what the implications would be.
        \item The authors should reflect on the scope of the claims made, e.g., if the approach was only tested on a few datasets or with a few runs. In general, empirical results often depend on implicit assumptions, which should be articulated.
        \item The authors should reflect on the factors that influence the performance of the approach. For example, a facial recognition algorithm may perform poorly when image resolution is low or images are taken in low lighting. Or a speech-to-text system might not be used reliably to provide closed captions for online lectures because it fails to handle technical jargon.
        \item The authors should discuss the computational efficiency of the proposed algorithms and how they scale with dataset size.
        \item If applicable, the authors should discuss possible limitations of their approach to address problems of privacy and fairness.
        \item While the authors might fear that complete honesty about limitations might be used by reviewers as grounds for rejection, a worse outcome might be that reviewers discover limitations that aren't acknowledged in the paper. The authors should use their best judgment and recognize that individual actions in favor of transparency play an important role in developing norms that preserve the integrity of the community. Reviewers will be specifically instructed to not penalize honesty concerning limitations.
    \end{itemize}

\item {\bf Theory assumptions and proofs}
    \item[] Question: For each theoretical result, does the paper provide the full set of assumptions and a complete (and correct) proof?
    \item[] Answer: \answerNA{} 
    \item[] Justification: We don't have theoretical result
    \item[] Guidelines:
    \begin{itemize}
        \item The answer NA means that the paper does not include theoretical results. 
        \item All the theorems, formulas, and proofs in the paper should be numbered and cross-referenced.
        \item All assumptions should be clearly stated or referenced in the statement of any theorems.
        \item The proofs can either appear in the main paper or the supplemental material, but if they appear in the supplemental material, the authors are encouraged to provide a short proof sketch to provide intuition. 
        \item Inversely, any informal proof provided in the core of the paper should be complemented by formal proofs provided in appendix or supplemental material.
        \item Theorems and Lemmas that the proof relies upon should be properly referenced. 
    \end{itemize}

    \item {\bf Experimental result reproducibility}
    \item[] Question: Does the paper fully disclose all the information needed to reproduce the main experimental results of the paper to the extent that it affects the main claims and/or conclusions of the paper (regardless of whether the code and data are provided or not)?
    \item[] Answer: \answerYes{} 
    \item[] Justification: Yes. The paper provides sufficient information to reproduce the main experimental results. We release the dataset and describe the experimental setup, intervention methods, model training procedures, and evaluation metrics in detail in the main paper, ensuring transparency and reproducibility of the core findings.
    \item[] Guidelines:
    \begin{itemize}
        \item The answer NA means that the paper does not include experiments.
        \item If the paper includes experiments, a No answer to this question will not be perceived well by the reviewers: Making the paper reproducible is important, regardless of whether the code and data are provided or not.
        \item If the contribution is a dataset and/or model, the authors should describe the steps taken to make their results reproducible or verifiable. 
        \item Depending on the contribution, reproducibility can be accomplished in various ways. For example, if the contribution is a novel architecture, describing the architecture fully might suffice, or if the contribution is a specific model and empirical evaluation, it may be necessary to either make it possible for others to replicate the model with the same dataset, or provide access to the model. In general. releasing code and data is often one good way to accomplish this, but reproducibility can also be provided via detailed instructions for how to replicate the results, access to a hosted model (e.g., in the case of a large language model), releasing of a model checkpoint, or other means that are appropriate to the research performed.
        \item While NeurIPS does not require releasing code, the conference does require all submissions to provide some reasonable avenue for reproducibility, which may depend on the nature of the contribution. For example
        \begin{enumerate}
            \item If the contribution is primarily a new algorithm, the paper should make it clear how to reproduce that algorithm.
            \item If the contribution is primarily a new model architecture, the paper should describe the architecture clearly and fully.
            \item If the contribution is a new model (e.g., a large language model), then there should either be a way to access this model for reproducing the results or a way to reproduce the model (e.g., with an open-source dataset or instructions for how to construct the dataset).
            \item We recognize that reproducibility may be tricky in some cases, in which case authors are welcome to describe the particular way they provide for reproducibility. In the case of closed-source models, it may be that access to the model is limited in some way (e.g., to registered users), but it should be possible for other researchers to have some path to reproducing or verifying the results.
        \end{enumerate}
    \end{itemize}

\item {\bf Open access to data and code}
    \item[] Question: Does the paper provide open access to the data and code, with sufficient instructions to faithfully reproduce the main experimental results, as described in supplemental material?
    \item[] Answer: \answerYes{} 
    \item[] Justification: We have released the complete GitHub repository with dataset and code.
    \item[] Guidelines:
    \begin{itemize}
        \item The answer NA means that paper does not include experiments requiring code.
        \item Please see the NeurIPS code and data submission guidelines (\url{https://nips.cc/public/guides/CodeSubmissionPolicy}) for more details.
        \item While we encourage the release of code and data, we understand that this might not be possible, so “No” is an acceptable answer. Papers cannot be rejected simply for not including code, unless this is central to the contribution (e.g., for a new open-source benchmark).
        \item The instructions should contain the exact command and environment needed to run to reproduce the results. See the NeurIPS code and data submission guidelines (\url{https://nips.cc/public/guides/CodeSubmissionPolicy}) for more details.
        \item The authors should provide instructions on data access and preparation, including how to access the raw data, preprocessed data, intermediate data, and generated data, etc.
        \item The authors should provide scripts to reproduce all experimental results for the new proposed method and baselines. If only a subset of experiments are reproducible, they should state which ones are omitted from the script and why.
        \item At submission time, to preserve anonymity, the authors should release anonymized versions (if applicable).
        \item Providing as much information as possible in supplemental material (appended to the paper) is recommended, but including URLs to data and code is permitted.
    \end{itemize}

\item {\bf Experimental setting/details}
    \item[] Question: Does the paper specify all the training and test details (e.g., data splits, hyperparameters, how they were chosen, type of optimizer, etc.) necessary to understand the results?
    \item[] Answer: \answerYes{} 
    \item[] Justification: Yes, we give a details about how we select data, and how we constructed our dataset. Also the training configuration.
    \item[] Guidelines:
    \begin{itemize}
        \item The answer NA means that the paper does not include experiments.
        \item The experimental setting should be presented in the core of the paper to a level of detail that is necessary to appreciate the results and make sense of them.
        \item The full details can be provided either with the code, in appendix, or as supplemental material.
    \end{itemize}

\item {\bf Experiment statistical significance}
    \item[] Question: Does the paper report error bars suitably and correctly defined or other appropriate information about the statistical significance of the experiments?
    \item[] Answer: \answerNo{} 
    \item[] Justification: We believe that our experiment does not require this.
    \item[] Guidelines:
    \begin{itemize}
        \item The answer NA means that the paper does not include experiments.
        \item The authors should answer "Yes" if the results are accompanied by error bars, confidence intervals, or statistical significance tests, at least for the experiments that support the main claims of the paper.
        \item The factors of variability that the error bars are capturing should be clearly stated (for example, train/test split, initialization, random drawing of some parameter, or overall run with given experimental conditions).
        \item The method for calculating the error bars should be explained (closed form formula, call to a library function, bootstrap, etc.)
        \item The assumptions made should be given (e.g., Normally distributed errors).
        \item It should be clear whether the error bar is the standard deviation or the standard error of the mean.
        \item It is OK to report 1-sigma error bars, but one should state it. The authors should preferably report a 2-sigma error bar than state that they have a 96\% CI, if the hypothesis of Normality of errors is not verified.
        \item For asymmetric distributions, the authors should be careful not to show in tables or figures symmetric error bars that would yield results that are out of range (e.g. negative error rates).
        \item If error bars are reported in tables or plots, The authors should explain in the text how they were calculated and reference the corresponding figures or tables in the text.
    \end{itemize}

\item {\bf Experiments compute resources}
    \item[] Question: For each experiment, does the paper provide sufficient information on the computer resources (type of compute workers, memory, time of execution) needed to reproduce the experiments?
    \item[] Answer: \answerNo{} 
    \item[] Justification: We are working with inference only, compute resources is not the factor of any of our experiments.
    \item[] Guidelines:
    \begin{itemize}
        \item The answer NA means that the paper does not include experiments.
        \item The paper should indicate the type of compute workers CPU or GPU, internal cluster, or cloud provider, including relevant memory and storage.
        \item The paper should provide the amount of compute required for each of the individual experimental runs as well as estimate the total compute. 
        \item The paper should disclose whether the full research project required more compute than the experiments reported in the paper (e.g., preliminary or failed experiments that didn't make it into the paper). 
    \end{itemize}
    
\item {\bf Code of ethics}
    \item[] Question: Does the research conducted in the paper conform, in every respect, with the NeurIPS Code of Ethics \url{https://neurips.cc/public/EthicsGuidelines}?
    \item[] Answer: \answerYes{} 
    \item[] Justification: We don't have any code of ethics issues in this paper
    \item[] Guidelines:
    \begin{itemize}
        \item The answer NA means that the authors have not reviewed the NeurIPS Code of Ethics.
        \item If the authors answer No, they should explain the special circumstances that require a deviation from the Code of Ethics.
        \item The authors should make sure to preserve anonymity (e.g., if there is a special consideration due to laws or regulations in their jurisdiction).
    \end{itemize}

\item {\bf Broader impacts}
    \item[] Question: Does the paper discuss both potential positive societal impacts and negative societal impacts of the work performed?
    \item[] Answer: \answerNA{} 
    \item[] Justification: We believe this is not related to our work.
    \item[] Guidelines:
    \begin{itemize}
        \item The answer NA means that there is no societal impact of the work performed.
        \item If the authors answer NA or No, they should explain why their work has no societal impact or why the paper does not address societal impact.
        \item Examples of negative societal impacts include potential malicious or unintended uses (e.g., disinformation, generating fake profiles, surveillance), fairness considerations (e.g., deployment of technologies that could make decisions that unfairly impact specific groups), privacy considerations, and security considerations.
        \item The conference expects that many papers will be foundational research and not tied to particular applications, let alone deployments. However, if there is a direct path to any negative applications, the authors should point it out. For example, it is legitimate to point out that an improvement in the quality of generative models could be used to generate deepfakes for disinformation. On the other hand, it is not needed to point out that a generic algorithm for optimizing neural networks could enable people to train models that generate Deepfakes faster.
        \item The authors should consider possible harms that could arise when the technology is being used as intended and functioning correctly, harms that could arise when the technology is being used as intended but gives incorrect results, and harms following from (intentional or unintentional) misuse of the technology.
        \item If there are negative societal impacts, the authors could also discuss possible mitigation strategies (e.g., gated release of models, providing defenses in addition to attacks, mechanisms for monitoring misuse, mechanisms to monitor how a system learns from feedback over time, improving the efficiency and accessibility of ML).
    \end{itemize}
    
\item {\bf Safeguards}
    \item[] Question: Does the paper describe safeguards that have been put in place for responsible release of data or models that have a high risk for misuse (e.g., pretrained language models, image generators, or scraped datasets)?
    \item[] Answer: \answerNA{} 
    \item[] Justification: All data we used are published public dataset.
    \item[] Guidelines:
    \begin{itemize}
        \item The answer NA means that the paper poses no such risks.
        \item Released models that have a high risk for misuse or dual-use should be released with necessary safeguards to allow for controlled use of the model, for example by requiring that users adhere to usage guidelines or restrictions to access the model or implementing safety filters. 
        \item Datasets that have been scraped from the Internet could pose safety risks. The authors should describe how they avoided releasing unsafe images.
        \item We recognize that providing effective safeguards is challenging, and many papers do not require this, but we encourage authors to take this into account and make a best faith effort.
    \end{itemize}

\item {\bf Licenses for existing assets}
    \item[] Question: Are the creators or original owners of assets (e.g., code, data, models), used in the paper, properly credited and are the license and terms of use explicitly mentioned and properly respected?
    \item[] Answer: \answerYes{} 
    \item[] Justification: We cited all models, and dataset we used in this paper.
    \item[] Guidelines:
    \begin{itemize}
        \item The answer NA means that the paper does not use existing assets.
        \item The authors should cite the original paper that produced the code package or dataset.
        \item The authors should state which version of the asset is used and, if possible, include a URL.
        \item The name of the license (e.g., CC-BY 4.0) should be included for each asset.
        \item For scraped data from a particular source (e.g., website), the copyright and terms of service of that source should be provided.
        \item If assets are released, the license, copyright information, and terms of use in the package should be provided. For popular datasets, \url{paperswithcode.com/datasets} has curated licenses for some datasets. Their licensing guide can help determine the license of a dataset.
        \item For existing datasets that are re-packaged, both the original license and the license of the derived asset (if it has changed) should be provided.
        \item If this information is not available online, the authors are encouraged to reach out to the asset's creators.
    \end{itemize}

\item {\bf New assets}
    \item[] Question: Are new assets introduced in the paper well documented and is the documentation provided alongside the assets?
    \item[] Answer: \answerYes{}
    \item[] Justification: We introduce a new annotated dataset to support cognitive function analysis, which is submitted in the supplementary materials. We also provide detailed documentation describing its construction process, structure, and usage guidelines in the paper.
    \item[] Guidelines:
    \begin{itemize}
        \item The answer NA means that the paper does not release new assets.
        \item Researchers should communicate the details of the dataset/code/model as part of their submissions via structured templates. This includes details about training, license, limitations, etc. 
        \item The paper should discuss whether and how consent was obtained from people whose asset is used.
        \item At submission time, remember to anonymize your assets (if applicable). You can either create an anonymized URL or include an anonymized zip file.
    \end{itemize}

\item {\bf Crowdsourcing and research with human subjects}
    \item[] Question: For crowdsourcing experiments and research with human subjects, does the paper include the full text of instructions given to participants and screenshots, if applicable, as well as details about compensation (if any)? 
    \item[] Answer: \answerNA{}
    \item[] Justification: Not realted
    \item[] Guidelines:
    \begin{itemize}
        \item The answer NA means that the paper does not involve crowdsourcing nor research with human subjects.
        \item Including this information in the supplemental material is fine, but if the main contribution of the paper involves human subjects, then as much detail as possible should be included in the main paper. 
        \item According to the NeurIPS Code of Ethics, workers involved in data collection, curation, or other labor should be paid at least the minimum wage in the country of the data collector. 
    \end{itemize}

\item {\bf Institutional review board (IRB) approvals or equivalent for research with human subjects}
    \item[] Question: Does the paper describe potential risks incurred by study participants, whether such risks were disclosed to the subjects, and whether Institutional Review Board (IRB) approvals (or an equivalent approval/review based on the requirements of your country or institution) were obtained?
    \item[] Answer: \answerNA{} 
    \item[] Justification: Not related
    \item[] Guidelines:
    \begin{itemize}
        \item The answer NA means that the paper does not involve crowdsourcing nor research with human subjects.
        \item Depending on the country in which research is conducted, IRB approval (or equivalent) may be required for any human subjects research. If you obtained IRB approval, you should clearly state this in the paper. 
        \item We recognize that the procedures for this may vary significantly between institutions and locations, and we expect authors to adhere to the NeurIPS Code of Ethics and the guidelines for their institution. 
        \item For initial submissions, do not include any information that would break anonymity (if applicable), such as the institution conducting the review.
    \end{itemize}

\item {\bf Declaration of LLM usage}
    \item[] Question: Does the paper describe the usage of LLMs if it is an important, original, or non-standard component of the core methods in this research? Note that if the LLM is used only for writing, editing, or formatting purposes and does not impact the core methodology, scientific rigorousness, or originality of the research, declaration is not required.
    \item[] Answer: \answerYes{} 
    \item[] Justification: We used LLMs as an integral part of our dataset construction process. Specifically, LLMs were used to generate intermediate reasoning steps and candidate answers under controlled prompting. All prompts, generation procedures, and filtering steps are clearly documented in the main paper and supplementary materials to ensure transparency and reproducibility. 
    \item[] Guidelines:
    \begin{itemize}
        \item The answer NA means that the core method development in this research does not involve LLMs as any important, original, or non-standard components.
        \item Please refer to our LLM policy (\url{https://neurips.cc/Conferences/2025/LLM}) for what should or should not be described.
    \end{itemize}

\end{enumerate}

\newpage
\appendix

\section{Appendix}

\subsection{The cognitive function distribution of other models}
\label{app-heatmaps}

\begin{figure}[H]
  \centering
  \includegraphics[width=0.9\linewidth]{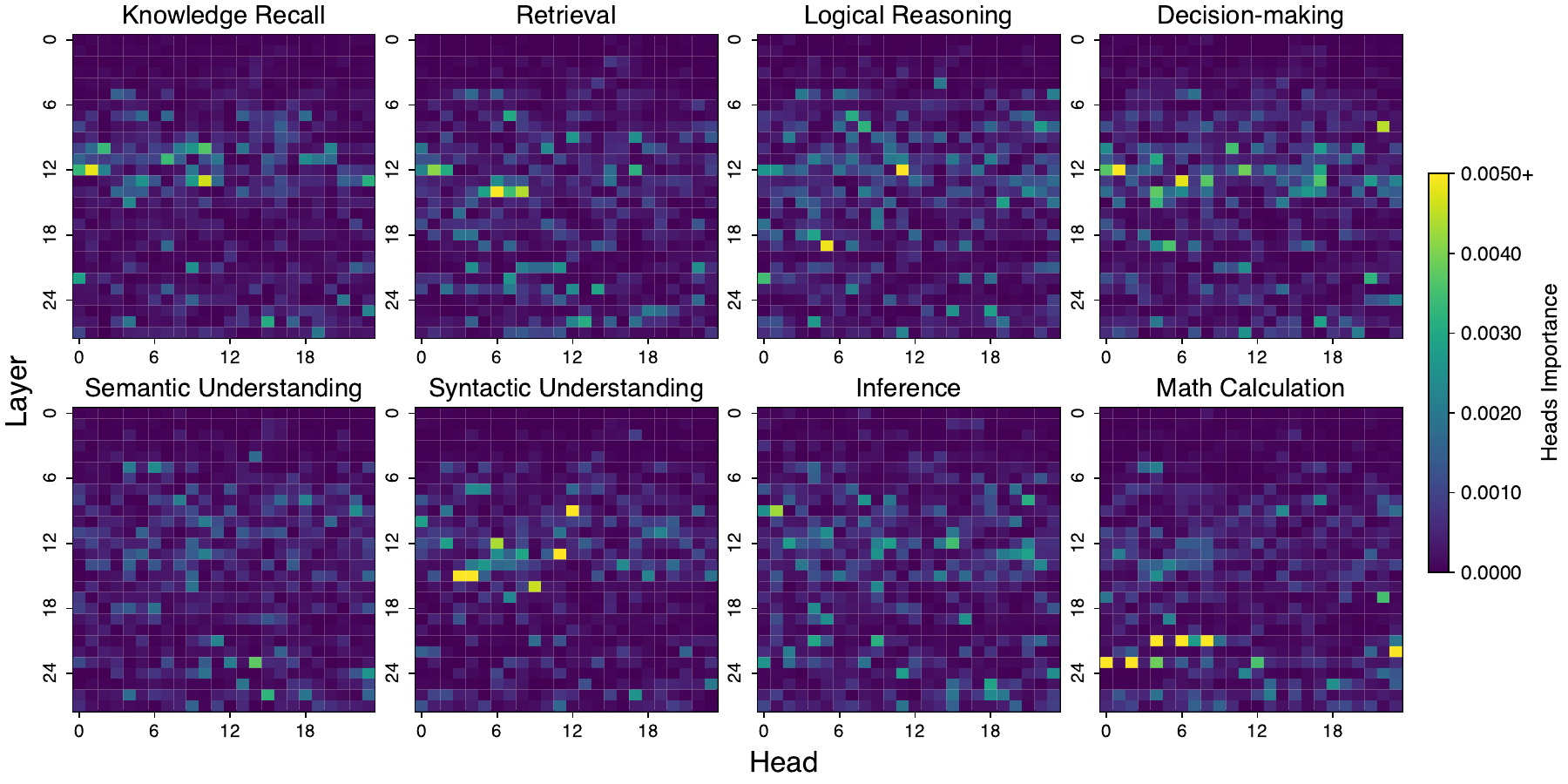} 
  \caption{The existence of cognitive heads in Llama3.2-3B-instruct responsible for eight distinct functions in complex reasoning tasks. The x-axis represents the head index, while the y-axis indicates the layer index.}
  \label{fig:heatmaps-llama33b}
\end{figure}

\begin{figure}[H]
  \centering
  \includegraphics[width=0.9\linewidth]{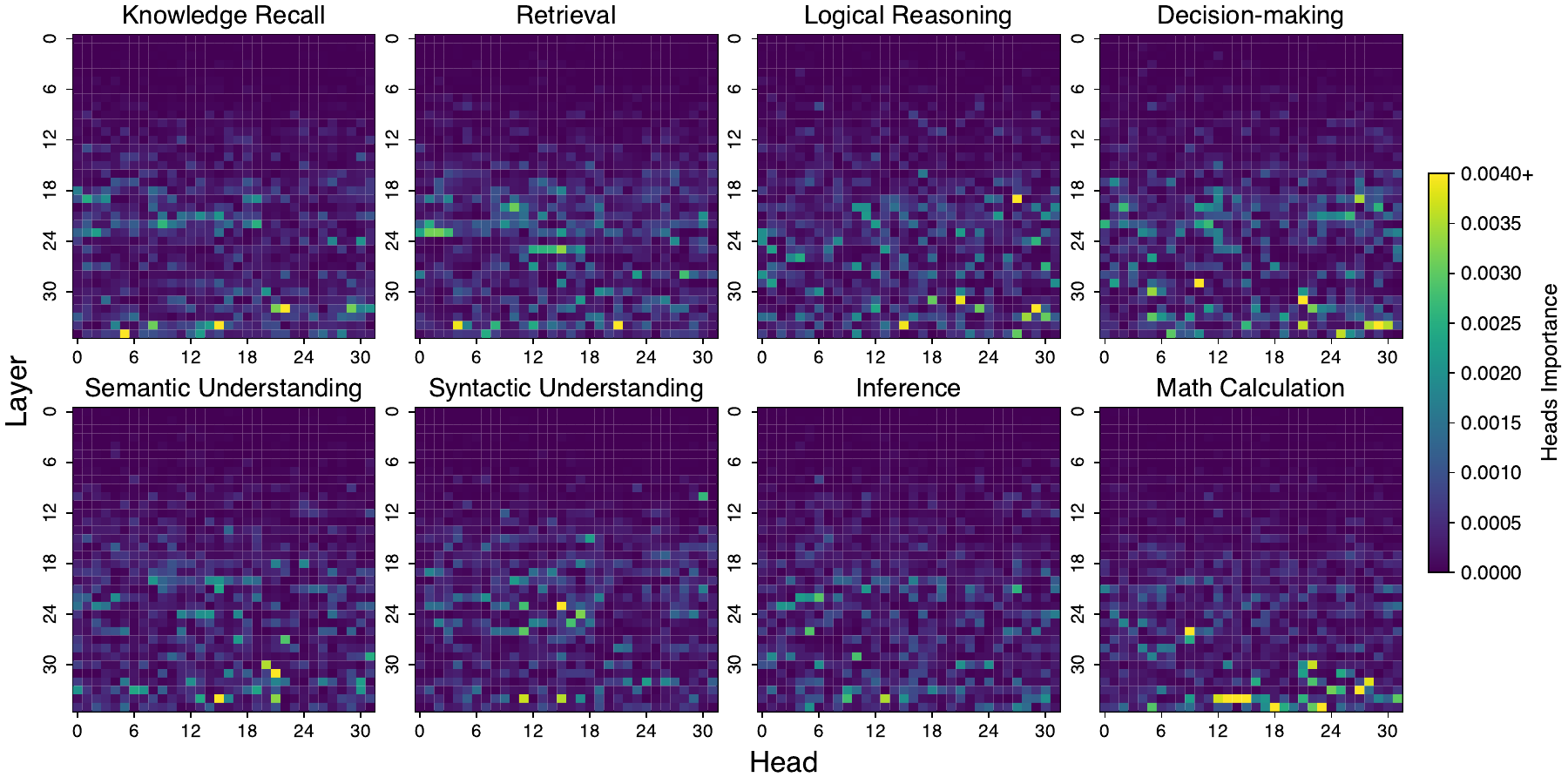} 
  \caption{The existence of cognitive heads in Qwen3-8B responsible for eight distinct functions in complex reasoning tasks. The x-axis represents the head index, while the y-axis indicates the layer index.}
  \label{fig:heatmaps-qwen38b}
\end{figure}

\begin{figure}[H]
  \centering
  \includegraphics[width=0.9\linewidth]{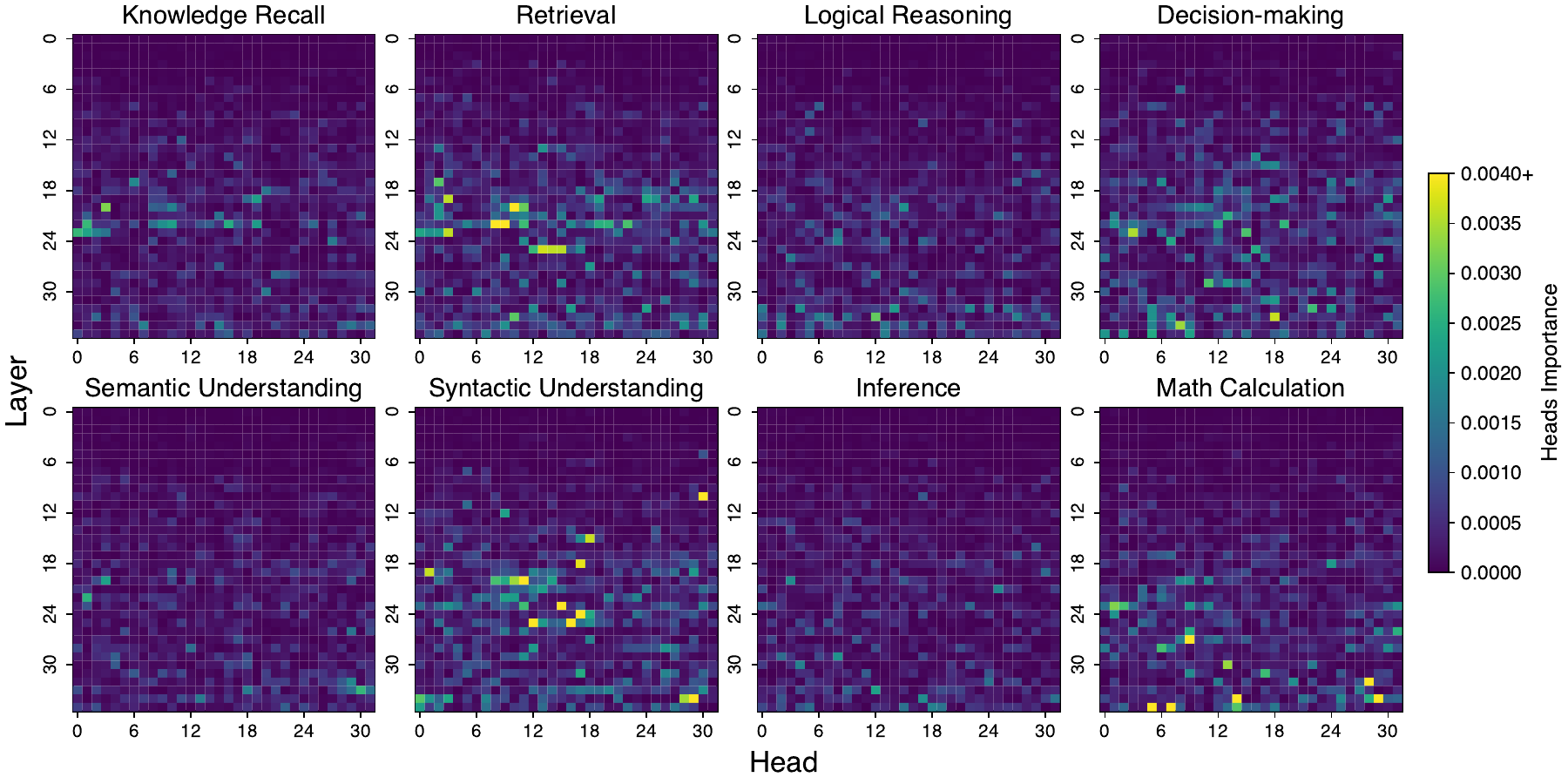} 
  \caption{The existence of cognitive heads in Qwen3-4B responsible for eight distinct functions in complex reasoning tasks. The x-axis represents the head index, while the y-axis indicates the layer index.}
  \label{fig:heatmaps-qwen34b}
\end{figure}

\begin{figure}[H]
  \centering
  \includegraphics[width=0.9\linewidth]{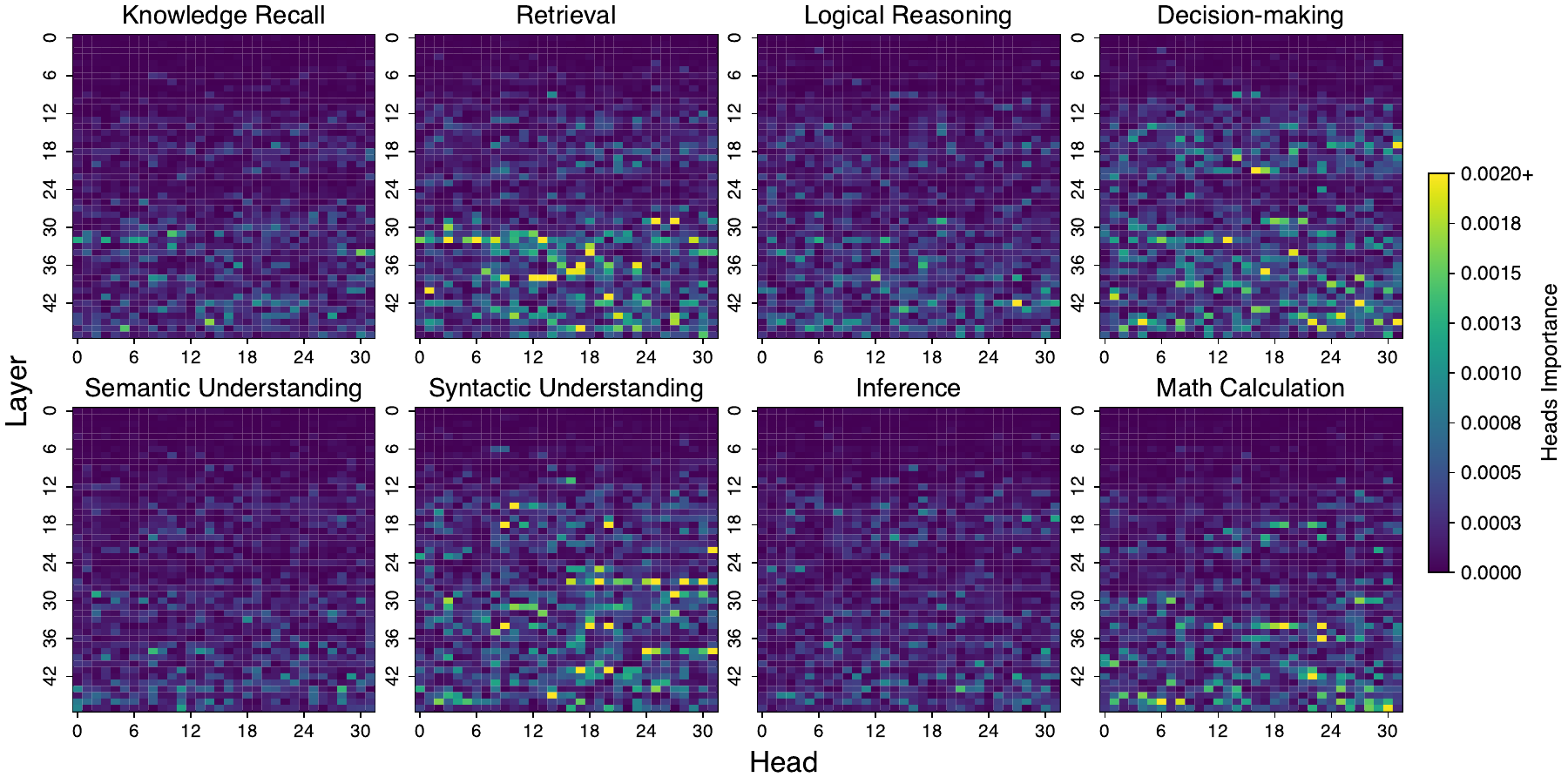} 
  \caption{The existence of cognitive heads in Yi-1.5-9B responsible for eight distinct functions in complex reasoning tasks. The x-axis represents the head index, while the y-axis indicates the layer index.}
  \label{fig:heatmaps-yi9b}
\end{figure}

\begin{figure}[]
  \centering
  \includegraphics[width=0.9\linewidth]{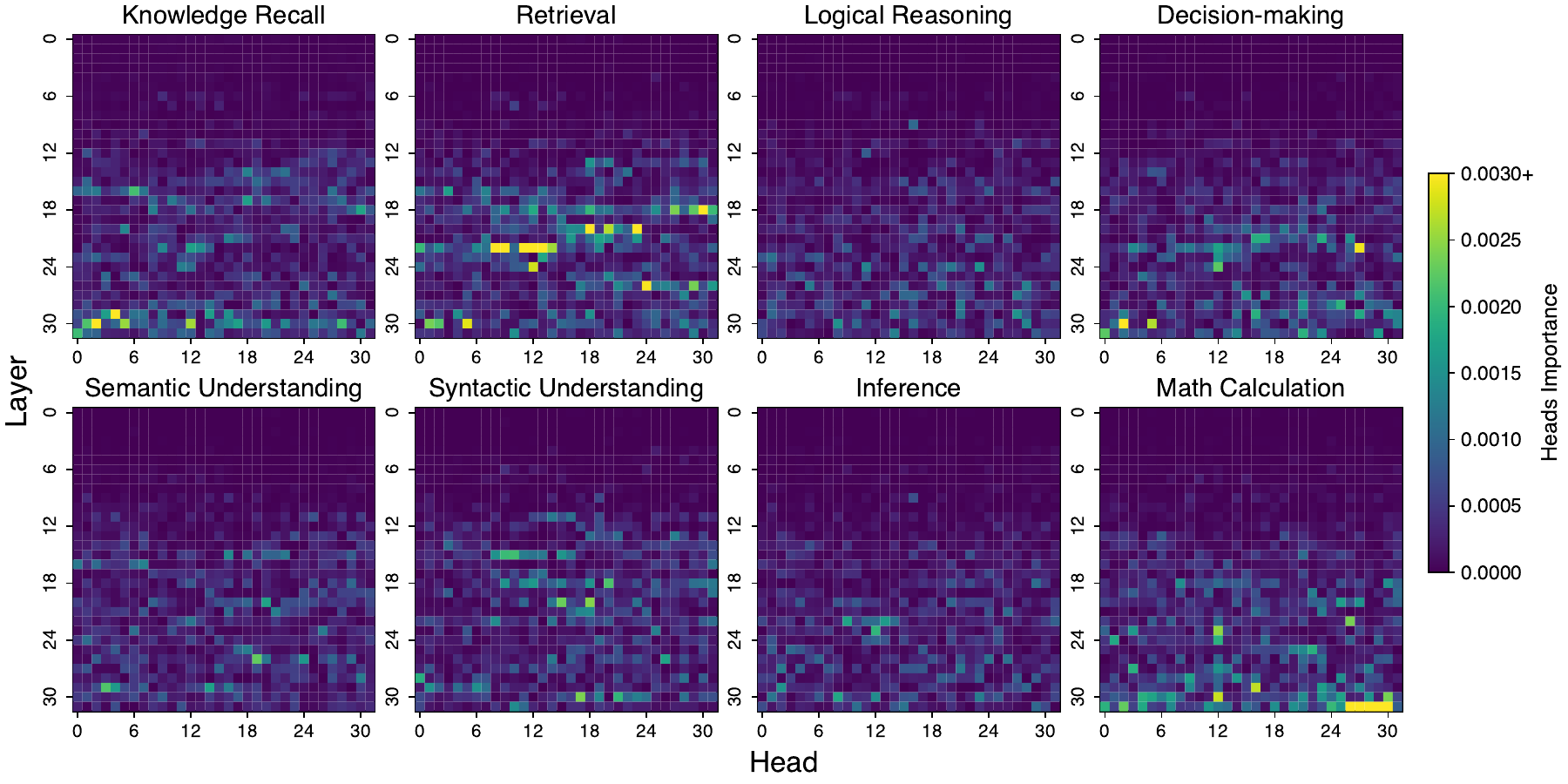} 
  \caption{The existence of cognitive heads in Yi-1.5-6B responsible for eight distinct functions in complex reasoning tasks. The x-axis represents the head index, while the y-axis indicates the layer index.}
  \label{fig:heatmaps-yi6b}
\end{figure}

We present the heatmaps for the remaining five models in this section. The results reveal a notable universality in the sparsity patterns of attention heads across different architectures. Moreover, models within the same family tend to exhibit similar sparsity distributions. For instance, Llama3.2-3B (Figure~\ref{fig:heatmaps-llama33b}) and Llama3.1-8B (Figure~\ref{fig:heatmaps}) share comparable patterns, as do Qwen3-4B (Figure~\ref{fig:heatmaps-qwen34b}) and Qwen3-8B (Figure~\ref{fig:heatmaps-qwen38b}), as well as Yi-1.5-6B (Figure~\ref{fig:heatmaps-yi6b}) and Yi-1.5-9B (Figure~\ref{fig:heatmaps-yi9b}). This consistency is likely due to the shared architectural design and similar pretraining data within each model family.

\subsection{Importance curve}
\label{app-curve}

\begin{figure}[H]
  \centering
  \includegraphics[width=0.8\linewidth]{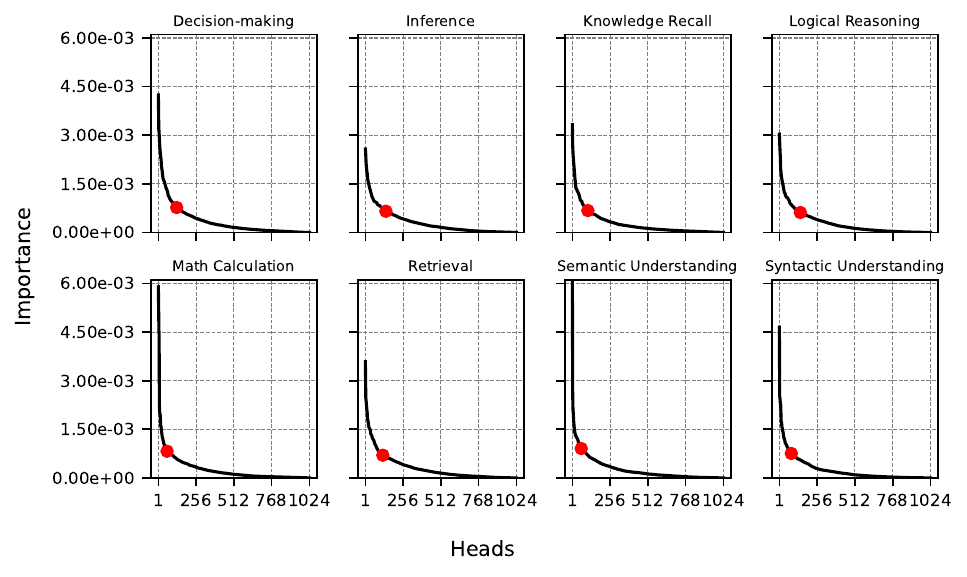} 
  \caption{Importance curve for eight functions, Llama3.1-8B-instruct.}
  \label{fig:elbow}
\end{figure}

We ranked the importance scores and identified the elbow point, as illustrated in Figure~\ref{fig:elbow}.

\subsection{MLP}\label{mlp}

We train a two-layer multi-class MLP for cognitive function classification. The first layer applies a shared linear projection to each multi-head representation vector, reducing each to a 64-dimensional embedding. These embeddings are then flattened and concatenated into a single vector of size $64\times number of heads$. This vector is fed into a hidden layer with 512 units, followed by a ReLU activation and a dropout with a rate of 0.3. The final output layer maps the 512-dimensional hidden representation to the set of cognitive function labels.

The model is trained using the Adam optimizer with a learning rate of $10^{-4}$ and a cross-entropy loss. Training proceeds for 100 epochs.
The test accuracy of our classification method across all LLM models is summarized in the Table \ref{app:test}. 

\begin{table}[]
\centering
\caption{The test accuracy (\%) of probing method on different LLMs.}
\resizebox{0.95\linewidth}{!}{
\begin{tabular}{lcccccc}
\toprule
 \textbf{Dataset} & \textbf{Llama3.1-8B-instruct} & \textbf{Llama3.2-3B-instruct} & \textbf{Qwen3-8B} & \textbf{Qwen3-4B} & \textbf{Yi-1.5-9B} & \textbf{Yi-1.5-6B} \\
\midrule
CogQA & 83.73 & 79.80 & 84.71 & 80.79 & 77.56 & 75.18\\
\bottomrule
\end{tabular}
}
\label{app:test}
\end{table}


\subsection{Prompt for Generating CogQA}
\label{app:prompt_generating}

\begin{promptbox}
\textbf{Prompt:} You are an expert in analytical logical reasoning. You will be given a question along with its chain-of-thought process.
Your task is to break the question down into subquestions based on the chain-of-thought process, ensuring that all necessary steps for solving the problem and constructing the logical chain are included to simulate critical thinking.

Decompose the Question: Identify and formulate the key subquestions required to solve the main question logically.
Fill in Missing Steps: Ensure that all essential reasoning steps are explicitly stated.

NOTE: The information of chain-of-thought cannot be used directly if it doesn't exist in main query.
Each subquestion should be derived solely from the main query and the preceding subquestion.
Answer the Subquestions: Provide clear, step-by-step solutions for each subquestion.
Annotate Cognitive Skills: Identify and label the specific cognitive abilities required to answer each subquestion.
If you believe other cognitive skills are relevant, you may also consider incorporating them.
You will be given predefined labels along with their descriptions.
Your goal is to enhance the logical reasoning process by making it explicit and structured.

<cognitive\_skills>
**Retrieval**: Refers to the process of fetching relevant information from input text, typically involving the extraction of specific words, phrases, or sentences directly from the original text.
**Knowledge Recall**: Involves the storage and recall of domain-specific knowledge, such as concepts from math, physics, biology, etc. This is typically the internal knowledge base of a language model. (Corresponding to the memory head)
**Semantic Understanding**: Refers to the ability to comprehend and extract meaning from text or symbols by recognizing relationships between words, phrases, and concepts. It goes beyond syntactic understanding by grasping context, intent, and underlying knowledge.
**Syntactic Understanding*: Involves the ability to analyze and interpret the grammatical structure of sentences, including the roles and relationships of words, phrases, and clauses within the language.
**Math Calculation**: Refers to the process of performing arithmetic or mathematical operations to obtain a result. It involves applying mathematical concepts, such as addition, subtraction, multiplication, division, and more complex operations (e.g., algebra, calculus), to solve problems or derive values from given inputs.
**Inference**: Involves drawing conclusions based on existing evidence or information. It follows logical rules to deduce new statements or decisions from given information.
**Logical Reasoning**: The process of drawing conclusions based on a set of premises, following established rules of logic, used to ensure that decisions of people are coherent, consistent, and based on sound principles.
**Decision-making**: The process of making a choice in a selection task based on previous information or analysis.
<cognitive\_skills>

Here is the question:
<question>
{question}
<question>

Here is the chain-of-thought:
<chain-of-thought>
{cot}
<chain-of-thought>

Note

 - Your task is to break the question down into detailed subquestions, ensuring each subquestion can be answered using only one specific cognitive skill.
 - You need to create a structured and explicit reasoning process that simulates critical thinking while maintaining clarity and precision.
 - The subquestion needs to be easy to answer and the answer needs to be concise
 - The information of chain-of-thought cannot be used directly if it doesn't exist in main query.
 - Each subquestion should be derived solely from the main query and the preceding subquestion.
 - You CAN NOT retrieval information from chain-of-thought, but you can retrieval from question.
 - Your output should be formatted as a list of JSON objects, where each object represents a subquestion, its answer, and the required cognitive skill.
 - You should use the most efficient logic to analyze the problem and minimize the number of subquestions.

Output format
[
  {
    "subquestion": "<Subquestion text>",
    "answer": "<Concise answer>",
    "cognitive\_skill": "<Assigned cognitive skill>"
  },
  {
    "subquestion": "<Subquestion text>",
    "answer": "<Concise answer>",
    "cognitive\_skill": "<Assigned cognitive skill>"
  }
]

Your answer:

\end{promptbox}
\subsection{Annotations}\label{app:annotation}

To ensure the quality and reliability of the decomposed subQAC triplets in the CogQA dataset, we design a rigorous multi-stage annotation pipeline, combining expert review and model-based verification. The goal is to verify the logical validity of subquestions, the correctness of their associated cognitive function labels, and the accuracy of the answers.

\paragraph{Stage 1: Validating Subquestion Decomposition}

In the first stage, we evaluate whether the generated subquestions are logically sound and align with natural human reasoning. For each QA pair, three expert annotators (with backgrounds in linguistics or cognitive science) independently assess the validity of each subquestion. A subquestion is marked \texttt{true} if it meaningfully contributes to answering the main question and follows a logical reasoning trajectory. Otherwise, it is marked \texttt{false}.

If a subquestion depends on prior information—such as the question text or the answer—from another subquestion, the subquestion order must reflect this dependency. While some subquestions can be answered in parallel and are order-independent, others have prerequisite relationships that require a specific sequence.
As the overall reasoning structure often forms a graph, where both sequential and parallel dependencies coexist. During LLM inference, we include the previous subquestions and their corresponding subanswers in the prompt as prior information. Thus, the critical factor is not the ordering alone, but whether the prompt provides the necessary context to answer the current subquestion accurately.

We apply the following filtering criteria:
\begin{itemize}[noitemsep,topsep=0pt]
    \item \textbf{AI-Human Agreement}: If any annotator considers fewer than 60\% of the subquestions valid, the entire QA decomposition is discarded.
    \item \textbf{Inter-Annotator Agreement}: A subquestion is deemed invalid if at least two annotators mark it as \texttt{false}. If over 40\% of the subquestions in a QA pair are invalid under this rule, the whole QA pair is removed.
\end{itemize}

This filtering ensures that the retained QA decompositions follow coherent, cognitively plausible reasoning chains.

\paragraph{Stage 2: Verifying Cognitive Function Labels}

In the second stage, annotators evaluate the correctness of the cognitive function label \(c_i\) assigned to each subQAC triplet \((q_i, a_i, c_i)\). Three annotators independently mark each label as \texttt{true} or \texttt{false}. When discrepancies occur, annotators collaboratively reassign the correct cognitive label to ensure alignment with the underlying mental operation.

This step ensures that the categorization of subquestions accurately reflects established distinctions between information retrieval, semantic understanding, logical reasoning, and other cognitive processes.

\paragraph{Stage 3: Answer Verification via Model and Human Review}

In the final stage, we verify the correctness of each answer \(a_i\) using both automated and manual procedures. We employ the o4-mini model~\cite{o4mini2024}, known for its logical reasoning capabilities, to re-evaluate GPT-4o-generated answers. If o4-mini disagrees with GPT-4o, it provides an alternative answer. A human annotator then compares both answers and resolves discrepancies by supplying the correct one when necessary. Given the generally objective nature of answers, only one annotator is required for this task.

\paragraph{Annotation Outcome}

Following this multi-stage process, we retain 570 validated QA pairs, yielding a total of 3,402 high-quality subQAC triplets. Notably, we augment certain cognitive functions to ensure balance across categories. As a result, the original 570 QA pairs were expanded to 720 (including some duplicates), with each duplicated pair potentially associated with distinct subquestions and cognitive functions.
\subsection{CogQA Example}\label{app:exmaple}

Table~\ref{tab:cogqa-example} presents illustrative examples from the CogQA dataset. The main question and its corresponding answer are taken from the original dataset. Based on an analysis of the main question, a sequence of sub-questions, their answers, and associated cognitive function labels are generated in order.

\begin{table}[t]
\caption{Two examples from the CogQA dataset showing a main question, its final answer, and a breakdown into subquestions with answers and their corresponding cognitive function labels.}
\label{tab:cogqa-example}
\centering
\small
\setlength{\tabcolsep}{6pt}
\renewcommand{\arraystretch}{1.2}

\begin{tabular}{p{0.2\linewidth} | p{0.75\linewidth}}
\toprule
\multicolumn{2}{l}{\textbf{Example 1:}}\\\toprule
\textbf{Main Question} & A one-year subscription to a newspaper is offered with a 45\% discount. How much does the discounted subscription cost if a subscription normally costs \$80? \\
\midrule
\textbf{Answer} & We calculate first the discount: 80 $\times$ 45 / 100 = \$36. So, the discounted subscription amounts to 80 – 36 = \$44. \\
\bottomrule
\end{tabular}

\begin{tabular}{p{0.6\linewidth} | p{0.118\linewidth} | p{0.2\linewidth}}
\toprule
\textbf{Subquestion} & \textbf{Answer} & \textbf{Cognitive Label} \\
\midrule
1. What is the normal cost of a one-year subscription to the newspaper? & \$80 & Retrieval \\
\midrule
2. What is the discount percentage offered on the subscription? & 45\% & Retrieval \\
\midrule
3. How much is the discount amount in dollars for the subscription? & \$36 & Math Calculation \\
\midrule
4. What is the cost of the subscription after applying the discount? & \$44 & Math Calculation \\
\bottomrule
\end{tabular}

\begin{tabular}{p{0.2\linewidth} | p{0.75\linewidth}}
\toprule
\multicolumn{2}{l}{\textbf{Example 2:}}\\\toprule
\textbf{Main Question} & What does every person talk out of? Options: - name - hide - mother and father - mouth - heart \\
\midrule
\textbf{Answer} & By mouth, talking is done. Every person talk out of mouth.\\
\bottomrule
\end{tabular}

\begin{tabular}{p{0.6\linewidth} | p{0.118\linewidth} | p{0.2\linewidth}}
\toprule
\textbf{Subquestion} & \textbf{Answer} & \textbf{Cognitive Label} \\
\midrule
1. What is the primary function of talking? & To communicate verbally. & Knowledge Recall \\

\midrule
2. Which part of the human body is primarily used for verbal communication? & Mouth & Knowledge Recall \\
\midrule
3. Based on the options provided, which option corresponds to the part used for verbal communication?
& Mouth & Decision-making \\

\bottomrule
\end{tabular}

\end{table}








\subsection{Prompt for Question Asking}\label{app:prompt_q}


\begin{promptbox}
\textbf{Prompt:} You are an expert in analytical and logical reasoning. You will be given a main question and prior knowledge in chain-of-thought (CoT) format.
Your task is to answer a follow-up subquestion using the information provided.

Here is the main question:

<main\_question>
{question}
</main\_question>

Here is the prior knowledge in chain-of-thought (CoT) format:

<prior\_knowledge>
{cot}
</prior\_knowledge>

Here is the subquestion:

<subquestion>
{subquestion}
</subquestion>

Instructions:

- Answer the subquestion carefully.

- You can use the information in the prior\_knowledge to help you answer the subquestion.

- Your response should be clear and concise.

- Stick to factual reasoning based on provided CoT.

- Do not include any explanation, commentary, or code.

- Do not output anything after the closing square bracket `]`.

Only output your final answer using this format:
[
    {{"answer": "<Your answer here>"}}
]

Your answer:
\end{promptbox}

\subsection{The number of cognitive heads for different LLMs}
\label{app-number}
The number of cognitive heads for each model is shown in Table~\ref{tab:heads_percent_table}.

\begin{table}[t]
\caption{Count (C) and percentage (\%) of attention heads exceeding elbow thresholds for each cognitive function across six models.}
\label{tab:heads_percent_table}
\centering
\small
\setlength{\tabcolsep}{2pt}
\begin{tabular}{l|cccccccccccccccc}
\toprule
\textbf{Model} & \multicolumn{2}{c}{\textbf{Recall}} & \multicolumn{2}{c}{\textbf{Retrieval}} & \multicolumn{2}{c}{\textbf{Logical}} & \multicolumn{2}{c}{\textbf{Decision}} & \multicolumn{2}{c}{\textbf{Semantic}} & \multicolumn{2}{c}{\textbf{Syntactic}} & \multicolumn{2}{c}{\textbf{Inference}} & \multicolumn{2}{c}{\textbf{Math}} \\
\cmidrule(lr){2-3}
\cmidrule(lr){4-5}
\cmidrule(lr){6-7}
\cmidrule(lr){8-9}
\cmidrule(lr){10-11}
\cmidrule(lr){12-13}
\cmidrule(lr){14-15}
\cmidrule(lr){16-17}
 & C & \% & C & \% & C & \% & C & \% & C & \% & C & \% & C & \% & C & \% \\
\midrule
Llama3.1-8B-instruct
& 105 & 10.3 & 118 & 11.5 & 142 & 13.9 & 124 & 12.1
& 60 & 5.9 & 81 & 7.9 & 139 & 13.6 & 59 & 5.8 \\
Llama3.2-3B-instruct
& 95 & 14.1 & 62 & 9.2 & 95 & 14.1 & 87 & 12.9
& 90 & 13.4 & 63 & 9.4 & 98 & 14.6 & 35 & 5.2 \\
Qwen3-8B
& 119 & 10.3 & 115 & 10.0 & 114 & 9.9 & 87 & 7.6
& 68 & 5.9 & 108 & 9.4 & 178 & 15.5 & 61 & 5.3 \\
Qwen3-4B
& 115 & 10.0 & 94 & 8.2 & 120 & 10.4 & 170 & 14.8
& 143 & 12.4 & 106 & 9.2 & 109 & 9.5 & 99 & 8.6 \\
Yi-1.5-9B
& 200 & 13.0 & 134 & 8.7 & 134 & 8.7 & 174 & 11.3
& 218 & 14.2 & 140 & 9.1 & 173 & 11.3 & 167 & 10.9 \\
Yi-1.5-6B
& 118 & 11.5 & 90 & 8.8 & 200 & 19.5 & 93 & 9.1
& 99 & 9.7 & 142 & 13.9 & 146 & 14.3 & 67 & 6.5 \\
\bottomrule
\end{tabular}

\end{table}

\subsection{Ablation study - Different position of head activation}\label{app:abl}
In the main experiments, we use the top-k generated tokens and average their multi-head attention vectors. We also explore alternative strategies for extracting representations, including using the first generated token, the last generated token, the first meaningful token, and the average of all generated tokens. The corresponding results are shown in Table~\ref{fig:abl}.

Here, first is the first token, last is the last token, meaning\_first is the first meaning token (excluding formatting), top-k is the top-k most semantically important tokens, full is all tokens in the answer.We observe that top-k token masking leads to the most significant performance drop when masking the top-30 identified heads, indicating higher precision in identifying retrieval-relevant heads. Interestingly, last, meaning\_first, full, and top-k show similar performance trends. This is because different tokens in the output contribute to answering the question, and as the number of masked cognitive heads increases, the influence of token using decreases. Additionally, for Retrieval, the full answer is usually meaningful, whereas others like Math Calculation require semantically meaningful tokens. Based on these results, we choose top-k as our final setting. 


\begin{table}[H]
\centering
\caption{Attention heads associated with cognitive functions are selected based on different token positions. Accuracy and COMET scores are evaluated after intervention; lower values indicate better outcomes.}
\resizebox{0.95\linewidth}{!}{
\begin{tabular}{lcccccc}
\toprule
 \textbf{Model} & \textbf{Head\_num} & \textbf{Token\_use} & \textbf{Retrieval(comet)} & \textbf{Retrieval(acc)} & \textbf{Math(comet)} & \textbf{Math(acc)} \\
\midrule
Llama3.1-8B & 30 & first & 90.51 & 83.53 & 91.13 & 73.13 \\
Llama3.1-8B & 30 & last & 86.86 & 81.76 & 90.04 & 68.66\\
Llama3.1-8B & 30 & meaning\_first & 88.13 & 79.41 & 89.72 & 68.66 \\
Llama3.1-8B & 30 & full & 73.93 & 47.06 & 89.92 & 69.15\\
Llama3.1-8B & 30 & top-k & \textbf{70.05} & \textbf{46.47} & 89.32 & 67.16\\
\hline
Llama3.1-8B & 50 & first & 93.28 & 89.41 & 94.46 & 89.57 \\
Llama3.1-8B & 50 & last & 64.39 & 41.18 & 92.05 & 70.15 \\
Llama3.1-8B & 50 & meaning\_first & 62.90 & 34.12 & 84.60 & 60.69\\
Llama3.1-8B & 50 & full & \textbf{46.20} & \textbf{11.76} & 89.01 & 78.11\\
Llama3.1-8B & 50 & top-k & 65.64 & 47.76 & 89.65 & 70.15\\

\bottomrule
\end{tabular}
}
\label{fig:abl}
\end{table}

\subsection{Examples of top-$k$ tokens}
\label{app:topk}

The selected tokens are intended to semantically represent the generated answer. Below are examples for different cognitive functions for Llama3.1-8B-instruct:

\begin{table}[H]
\centering
\caption{Examples of question decomposition with cognitive functions and token selection.}
\begin{tabular}{p{3.5cm} p{2.5cm} p{2.5cm} p{2cm} p{2cm}}
\hline
\textbf{Main Question} & \textbf{Subquestion} & \textbf{Cognitive Function} & \textbf{Answer} & \textbf{Selected Tokens} \\
\hline

Given the sentence "A surfboarder catches the waves." can we conclude that "A surfboarder in the water."? (Options: yes / it is not possible to tell / no) &
What is typically required for a surfboarder to catch waves? &
Knowledge Recall &
The surfboarder needs to be in the water. &
\texttt{['surfboarder', 'needs', 'be', 'in', 'water']} \\
\hline
Is the following a factual statement? "Due to its high density, countries around the world use Palladium to mint coins." (Options: yes / no) &
What is the statement in question? &
Retrieval &
The statement in question is: Due to its high density, countries around the world use Palladium to mint coins. &
\texttt{['high', 'density', 'Palladium', 'mint', 'coins']} \\
\hline
A one-year subscription to a newspaper is offered with a 45\% discount. How much does the discounted subscription cost if a subscription normally costs \$80? &
How much is the discount amount in dollars for the subscription? &
Math Calculation &
36 &
\texttt{['36']} \\
\hline
\end{tabular}
\end{table}







We can see that the selected tokens semantically represent the answer. Note that we use all tokens when the number of tokens is fewer than 5.

\end{document}